\documentclass[twocolumn,prb,preprintnumbers,amsmath,amssymb]{aastex631} 
\usepackage{amsmath} 
\usepackage{amsfonts} 
\usepackage{comment}
\usepackage{graphicx}
\usepackage{dcolumn}
\usepackage{bm}
\usepackage{rotating}
\usepackage{xcolor}
\usepackage{color,soul}
\sethlcolor{white}

\begin{document}
\title{Callisto's Non-Resonant Orbit as an Outcome of Circum-Jovian Disk Substructure}
\author{Teng Ee Yap}
\author{Konstantin Batygin}
\affil{Division of Geological and Planetary Sciences, California Institute of Technology, Pasadena, CA 91125, USA.}

\begin{abstract}
\indent The Galilean moons of Io, Europa, and Ganymede exhibit a 4:2:1 commensurability in their mean motions, a configuration known as the Laplace resonance. The prevailing view for the origin of this three-body resonance involves the convergent migration of the moons, resulting from gas-driven torques in the circum-Jovian disk wherein they accreted. To account for Callisto's exclusion from the resonant chain, a late and/or slow accretion of the fourth and outermost Galilean moon is typically invoked, stalling its migration. Here, we consider an alternative scenario in which Callisto's non-resonant orbit is \textit{a consequence of disk substructure}. Using a suite of N-body simulations that self-consistently account for satellite-disk interactions, we show that a pressure bump can function as a migration trap, isolating Callisto and alleviating constraints on its timing of accretion. Our simulations position the bump interior to the birthplaces of all four moons. In exploring the impact of bump structure on simulation outcomes, we find that it cannot be too sharp nor flat to yield the observed orbital architecture. In particular, a ``goldilocks" zone is mapped in parameter space, corresponding to a well-defined range in bump aspect ratio. Within this range, Io, Europa, and Ganymede are sequentially trapped at the bump, and ushered across it through resonant lockstep migration with their neighboring, exterior moon. The implications of our work are discussed in the context of uncertainties regarding Callisto's interior structure, arising from the possibility of non-hydrostatic contributions to its shape and gravity field, unresolved by the \textit{Galileo} spacecraft. \\

\textit{Unified Astronomy Thesaurus concepts:} Galilean satellites (627), Planetary-disk interactions (2204), Planetary migration (2206).
\end{abstract}

\section{Introduction}
\subsection{Origin of the Laplace Resonance}
\indent Ever since their serendipitous discovery by Galileo in 1610, the Jovian moons of Io, Europa, Ganymede, and Callisto, have been captivating targets for comparative planetology. A centerpiece of any discussion on the origin and evolution of the Jovian system is the resonant configuration between the former three moons, first quantified by Laplace in the late 18th century. The eponymous, three-body resonance consists of two 2:1 mean motion commensurabilities. Today, tidal heating in these moons is attributed to eccentricity forcing through this resonance, fueling Io’s extensive volcanism \textcolor{blue}{(\textit{e.g.,} Peale et al., 1979; McEwen et al., 2000)}, and preserving subsurface oceans in Europa and Ganymede \textcolor{blue}{(\textit{e.g.,} Cassen et al., 1979; Carr et al., 1998; Kivelson et al., 2000, 2002)}. \\
\indent The stability of the Laplace resonance can be expressed via librating resonant arguments (i.e., the conjunction longitude of each resonant pair, in a frame co-moving with one of the two periapses). Henceforth denoting Io, Europa, and Ganymede as \textit{I}, \textit{E}, and \textit{G}, the arguments are given by
\begin{equation}
\begin{split}
&\theta_1 = \lambda_I - 2\lambda_E + \varpi_I \sim 0;\\
& \theta_2 =\lambda_I - 2\lambda_E + \varpi_E \sim \pi;\\
& \theta_3 =\lambda_E - 2\lambda_G + \varpi_E \sim 0,
\end{split}
\end{equation}
where $\lambda_X$ and $\varpi_X$ are the mean longitude and longitude of periapsis of moon \textit{X}, respectively. As such, \textit{I-E} conjunctions occur at the periapsis of \textit{I} but the apoapsis of \textit{E}. While \textit{E-G} conjunctions occur at the periapsis of \textit{E}, \textit{G} can be anywhere in its orbit (\textit{i.e.,} $\lambda_E - 2\lambda_G + \varpi_G$ circulates; \textcolor{blue}{Peale, 1999; Peale \& Lee, 2002}). Combining the last two relations, we arrive at the Laplace relation: $\theta_L = \lambda_I - 3\lambda_E + 2\lambda_G \sim \pi$, describing the 4:2:1 resonance chain. \\
\indent Models for the formation of the Laplace resonance fall into two major camps, invoking either (i) differential tidal expansion of orbits from tidal torques exerted by Jupiter \textcolor{blue}{(\textit{e.g.}, Goldreich \& Sciama, 1965; Yoder, 1979; Yoder \& Peale, 1981; Greenberg, 1987; Malhotra, 1991; Showman \& Malhotra, 1997)}, or (ii) convergent, disk-driven migration in the Jovian circumplanetary disk (henceforth the ``circum-Jovian" disk) within which the moons accreted \textcolor{blue}{(\textit{e.g.}, Canup \& Ward, 2002; Peale \& Lee, 2002; Sasaki et al., 2010; Madeira et al., 2021; Shibaike et al., 2019; Batygin \& Morbidelli, 2020)}. Notably, these two paradigms differ vastly in their age estimates for the said resonance. While the former generally constrains this age to $\ll$ 3 Gyr (the exact value dependent on Jupiter’s tidal quality factor $Q_J$), the latter implies an age coinciding with the lifetime of the circumsolar disk, which dissipated $\sim$4 Myr following the condensation of the Solar System’s (SS) first solids (i.e., Calcium-Aluminum-rich Inclusions; CAIs) $\sim$ 4.57 Gyr ago \textcolor{blue}{(Wang et al., 2017; Borlina et al., 2022; Weiss et al., 2021)}. In broad strokes, the tidal origin story posits that \textit{I} and \textit{E} were driven outwards by the dissipative tide raised on Jupiter. By virtue of its larger mass, \textit{I} approaches and eventually captures \textit{E} into the 2:1 mean motion resonance. Moving out in lock-step, the pair subsequently encounters the 4:2:1 resonance with Ganymede. Resonant capture is thus envisioned to occur from ``inside out," with ensuing forced eccentricities damped by tidal dissipation in the moons, mainly \textit{I}. \\
\indent The disk-driven scenario envisions resonant capture from ``outside in," whereby the moons converge upon the 2:1 commensurabilities via so-called ``Type-I" migration \textcolor{blue}{(Ward, 1997; Tanaka et al., 2002; Kley \& Nelson, 2012; Armitage, 2020)}. It has long been recognized that gravitational forces exerted by a planetary body on the disk material in which it is embedded lead to the launching of density waves at Lindblad resonances (\textit{i.e.}, mean motion resonances between the body and disk gas; \textcolor{blue}{Goldreich \& Tremaine, 1979, 1980}). Such waves, along with co-orbiting gas executing horseshoe orbits (in the frame of the body), exert torques on the body that generally (\textit{i.e.,} for most of the disk, wherein density falls with radial distance from the host star/planet) lead to inward migration. Accordingly, \textit{I} migrates inward until it encounters, and parks at, the disk inner edge (\textit{i.e.,} magnetospheric truncation radius; \textcolor{blue}{Ghosh \& Lamb, 1979; Ostriker \& Shu, 1995; Masset et al., 2006; Mohanty \& Shu, 2008}). Subsequently, \textit{E} and \textit{G} migrate towards \textit{I} and establish the Laplace resonance, with the timing of resonant captures dependent on their formation locations and disk structure. \\
\indent A primordial origin by convergent migration is favored on several grounds. For one, the tidal scenario requires approximately in situ accretion of the moons. Considering the substantial presence of water-ice in \textit{E}, \textit{G}, and Callisto (denoted \textit{C}; \textcolor{blue}{Kuskov \& Kronrod, 2001; Sohl et al., 2002}), this imposes an unrealistic constraint on models for the circum-Jovian disk: its midplane temperatures must be sufficiently low for the building blocks of the moons to lie beyond the water-ice sublimation front (henceforth denoted the ``ice-line"), despite their proximity to proto-Jupiter. Assuming (conservatively) an optically thin disk heated \textit{only} by passive irradiation from proto-Jupiter (\textit{i.e.,} neglecting viscous heating, which is expected to dominate the inner disk region of interest; \textcolor{blue}{\textit{e.g.,} Chambers, 2009; Batygin \& Morbidelli, 2020}), with an effective temperature of $T_J \sim 1400$ K, the ice-line would have resided at a jovicentric distance $\sim R_{J,Pr} (T_J^2/T_{ice}^2)$, where $T_{ice}\sim 170$ K represents the temperature of water-ice sublimation/condensation under nebular pressures (\textit{i.e.,} $\lesssim 10^{-3}$ bar), and $R_{J,Pr}$ Jupiter's primordial radius, estimated to be $\sim 2-2.5$ times its present-day radius $R_J \simeq 7 \times 10^7$m \textcolor{blue}{(Batygin \& Adams, 2025)}. This evaluates to $\sim 135-170 R_J$, well beyond the present-day location of \textit{C} at $\sim 26R_J$. \\
\indent Moving on, recent measurements of the mass-dependent S and Cl isotopic composition in \textit{I}'s atmosphere via the Atacama Large Millimeter/Submillimeter Array (ALMA) point to extensive volcanic activity and associated outgassing driven by tidal heating across most of the moon's lifetime  \textcolor{blue}{(de Kleer et al., 2024)}, thereby supporting a primordial Laplace resonance. Moreover, over the past decade, the discovery of numerous exoplanetary systems hosting planets in compact resonant chains (\textit{e.g.}, TRAPPIST-1, G\textcolor{blue}{Gillon et al., 2017; Luger et al., 2017; Pichierri et al., 2024a}; Kepler-223, \textcolor{blue}{Mills et al., 2016}) suggests convergent migration into resonance is a common process in the evolution of planetary systems, and as such, renders the disk-driven scenario a natural, and thus expected outcome. This is supported by the prevalent ``peas-in-a-pod" architecture (\textit{i.e.,} intra-system uniformity in size and mutual spacing) exhibited by systems that define the \textit{Kepler} survey \textcolor{blue}{(Weiss et al., 2018)}. This architecture naturally emerges from the ``breaking" of resonant configurations established prior to disk dissipation \textcolor{blue}{(\textit{e.g.,} Batygin \& Adams, 2017; Goldberg \& Batygin, 2022)}, a point buttressed by the observation that resonance chains are ubiquitous in young ($<$ 100 Myr) systems and decay with age on a timescale on the order of $\sim$ 100 Myr \textcolor{blue}{(Dai et al., 2024)}.

\subsection{Callisto's Non-Resonant Orbit}
\indent Any model for the creation of the Laplace resonance must address the exclusion of \textit{C} from the resonant chain. In particular, disk-driven scenarios must contend with the relatively large mass of \textit{C} (comparable to \textit{G}), which should have led to rapid inward migration. The prevailing explanation for \textit{C}'s non-resonant orbit is that it accreted late and/or slowly (\textit{i.e.,} $\gg$ 100 kyr), such that its migration was too slow to result in resonant capture with \textit{G} prior to disk dissipation \textcolor{blue}{(Peale \& Lee, 2002; Batygin \& Morbidelli, 2020)}. This is indirectly supported by gravity measurements from the Galileo spacecraft suggesting its interior is partially differentiated, characterized by a mixture of rock and ice extending from its center to beneath its outer ice shell \textcolor{blue}{(Anderson et al., 1998; Anderson et al., 2001)}. That is, \textit{C} formed too late for substantial interior heating by the short-lived radionuclide $^{26}$Al ($t_{1/2}\sim0.717$ Myr) to occur and/or too slowly (accretion timescale $\gtrsim$ 0.5 Myr) for sufficient retention of accretionary energy, leading to an interior too cold for extensive melting of water-ice \textcolor{blue}{(Schubert et al., 2004; Barr \& Canup, 2008)}. \\
\indent In several works, \textit{C}'s non-resonant orbit is described as a consequence of dynamical tides \textcolor{blue}{(Fuller et al., 2016)}, driving outward migration following the establishment of an 8:4:2:1 resonance chain \textcolor{blue}{(Shibaike et al., 2019; Madeira et al., 2021)}. That is, \textit{C} is originally locked in a 2:1 resonance with \textit{G}, but breaks free therefrom after disk dissipation. While conceivable, these models have yet to be validated by astrometric observations of Callisto's tidal migration timescale, and self-consistent simulations of Jupiter-satellite tidal dissipation effects remain absent. \\
\indent Disk substructure offers an alternative explanation for \textit{C}'s non-resonant orbit, as planet/satellite migration rates and directions depend strongly on local disk conditions, namely the gradient in gas density \textcolor{blue}{(\textit{e.g.,} Kley \& Nelson, 2012)}. Over the past decade, ALMA observations have revealed the ubiquity of axisymmetric (\textit{e.g.,} concentric dust rings) and non-axisymmetric (\textit{e.g.,} azimuthal dust trapping in vortices) substructures in protoplanetary disks \textcolor{blue}{(\textit{e.g.}, Flock et al., 2015; Andrews et al., 2018; Dullemond et al., 2018; see Birnstiel, 2024 for review)}, showing them to be incompatible with long-adopted models for smooth, power-law disks. Pressure bumps serving as dust traps notably lend themselves to explaining population-level properties of disks (\textit{e.g.,} the size-luminosity relation; \textcolor{blue}{Tripathi et al. 2017}) and exoplanetary systems (\textit{i.e.,} the intra-system uniformity in super-Earths; \textcolor{blue}{Batygin \& Morbidelli, 2023}). On the Cosmochemistry front, non-uniformity in disk structure is supported by the salient dichotomy in nucleosynthetic isotope anomalies between non-carbonaceous and carbonaceous SS materials, calling for a prolonged ($\gtrsim 4$ Myr) separation of dust reservoirs in the circumsolar disk (\textcolor{blue}{Warren, 2011; Burkhardt et al., 2019; Kleine et al., 2020; Yap \& Tissot, 2023; Tissot et al., 2025)}. This separation is typically attributed to either Jupiter's early formation (\textit{i.e.,} $\sim 20$ Earth masses within $1$ Myr from CAIs; \textcolor{blue}{Kruijer et al., 2017}), a pressure bump near Jupiter's formation region \textcolor{blue}{(Brasser \& Mojzsis, 2020)}, or preferential planetesimal formation at the silicate and water-ice sublimation fronts \textcolor{blue}{(Cuzzi \& Zahnle, 2004; Kretke \& Lin, 2007; Brauer et al., 2008; Ros \& Johansen, 2013; Drażkowska \& Alibert, 2017; Lichtenberg et al., 2021; Morbidelli et al., 2022)}. \\
\indent The prevalence of pressure bumps in protoplanetary disks provides confidence that they similarly occur in circumplanetary disks. Here, we show that such a bump in the circum-Jovian disk can serve as a migration “trap” for \textit{C}, keeping it isolated from \textit{I}, \textit{E}, and \textit{G} as they establish the Laplace resonance by convergent migration interior to the bump, thereby relaxing the need for its late/slow formation. The latter moons can form beyond the bump, as they are readily ``pushed" across it once captured into a temporary mean motion resonance with their neighboring exterior moon (\textit{i.e.,} the combined Type-I torque is sufficient to drive the interior moon over the bump). Accordingly, \textit{E} pushes \textit{I} across, \textit{G} pushes \textit{E}, and \textit{C} pushes \textit{G}. While we remain agnostic to the underlying origin for the bump invoked, we note that the ice-line serves as a natural place for its development \textcolor{blue}{(\textit{e.g.}, Kretke \& Lin, 2007; Brauer et al., 2008; Bitsch et al., 2015; Charnoz et al., 2021; Müller et al., 2021)}. \\
\indent The paper is structured as follows: In \textbf{Section 2}, we provide an overview of the circum-Jovian disk model adopted, as well as the disk-dependent parametrizations for Type-I migration and eccentricity damping. Derivations of key equations introduced in this section are relegated to the \textbf{Appendix}. A description of our simulations and their setup (\textit{e.g.,} initial conditions) is given in \textbf{Section 3}. In \textbf{Section 4}, we present our results, including a thorough analysis of a fiducial case and an exploration of how the structure of the pressure bump (\textit{i.e.}, its height and width) impacts the emergent architecture of the system. We discuss possible origins for pressure bumps, as well as the impact on our results from relaxing simplifying assumptions in our disk model in \textbf{Section 5}. There, we also consider variations on our envisioned scenario for the assembly of the Laplace resonance, and discuss our work in the context of Callisto's interior structure. Final remarks are given in \textbf{Section 6}. 
\begin{figure*} 
\centering
\scalebox{0.82}{\includegraphics{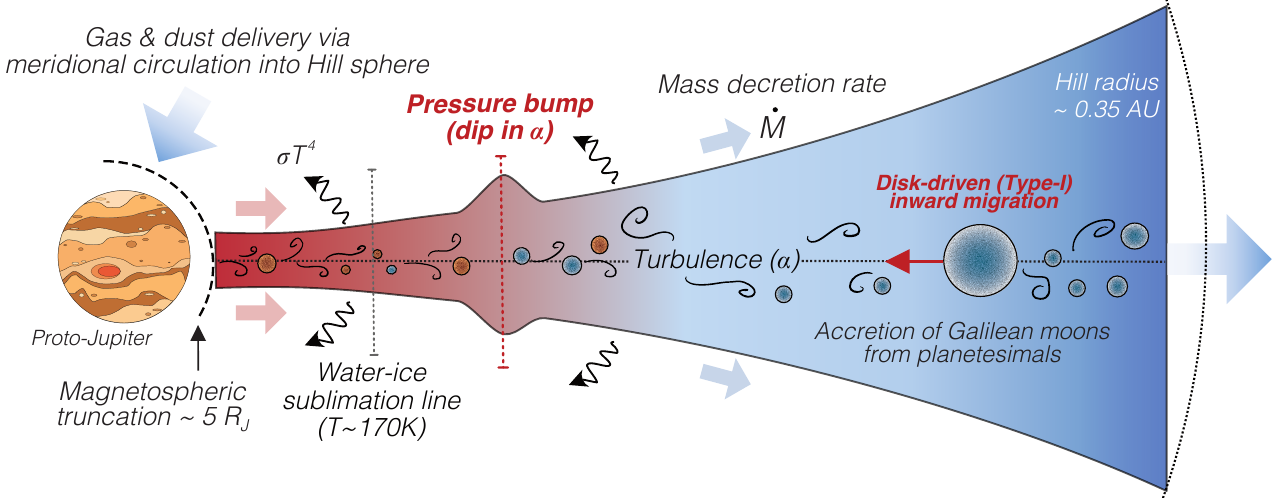}}
\caption{\textbf{Schematic of the circum-Jovian decretion disk}. Gas and dust from the circumsolar accretion disk are subsumed into the Jovian disk from approximately one hydrostatic scale height via meridional flows, and move outward beyond the magnetospheric truncation radius. Decretion $\dot{M}$ is driven by turbulence manifesting as a macroscopic viscosity, and parametrized by the Shakura-Sunyaev $\alpha$ parameter. The four Galilean moons are envisioned to form beyond the ice-line and pressure bump, and undergo Type-I migration inwards. The bump serves as a migration trap, preventing Callisto from convergent migration into resonance with Io, Europa, and Ganymede.}
\label{fig:Figure 1}
\end{figure*}
\section{Disk Model \& Type-I Forces}
\subsection{The Circum-Jovian Decretion Disk}
\indent Over the past decade, both hydrodynamical simulations \textcolor{blue}{(Tanigawa et al., 2012; Morbidelli et al., 2014; Szulágyi et al., 2022)} and direct observations \textcolor{blue}{(Teague et al., 2018, 2019)} of gas flow in the vicinity of gap-carving giant planets have inspired a dramatic re-imagination of circumplanetary disk formation and evolution. Unlike the circum\textit{stellar} disks in which they are hosted, circumplanetary disks are not mainly (\textit{i.e.,} across most of their radial extents) thought to be accreting, but \textit{decreting}. Indeed, the said studies suggest that gas and dust in such disks are vertically delivered onto the giant planet Hill sphere from approximately a hydrostatic scale height above the circumstellar disk midplane via meridional flows, resulting in decretion beyond the centrifugal radius (see \textbf{Fig. 1}). Following the approach of \textcolor{blue}{Batygin \& Morbidelli (2020)}, we adopt a steady-state viscous model for the circum-Jovian decretion disk. This model departs from the classic accretion scenario of \textcolor{blue}{Canup \& Ward (2002)}, but retains the key feature of being ``gas-starved." That is, gas is introduced into, and cycled out of, the disk throughout its lifetime, gradually providing the dust that will constitute the moons.\\
\indent The disk surface density profile $\Sigma(r)$ ($r$ being the radial jovicentric distance) serves as the backdrop to our simulations, to which the positions of the Galilean moons at each time step are mapped for the calculation of their respective Type-I migration (\textit{i.e.,} semi-major axis damping) and eccentricity damping rates (\textit{i.e.,} $\dot{a}$ and $\dot{e}$). Here, we outline the construction of $\Sigma(r)$, showing how a pressure (\textit{i.e.,} $\Sigma$) bump in the disk follows directly from the implementation of a Gaussian dip in an otherwise flat profile of the Shakura-Sunyaev $\alpha$ parameter for turbulent viscosity \textcolor{blue}{(Shakura \& Sunyaev, 1973)}. A detailed derivation of our disk model is provided in the \textbf{Appendix}.\\
\indent By conservation of mass and angular momentum, $\Sigma(r)$ takes the form \textcolor{blue}{(Lynden-Bell \& Pringle, 1974)}
\begin{equation}
\Sigma(r) = \frac{\dot{M}}{3\pi\nu}\left( \sqrt{{\frac{R_H}{r}}}-1\right),
\end{equation}
where $\dot{M}\sim r^0$ is the mass decretion rate, and $\nu$ the turbulent viscosity facilitating decretion. The former is taken to be $\sim 0.1 M_J/Myr$, with $M_J \simeq 1.9 \times 10^{27}$ kg being Jupiter's present-day mass. The latter is given by 
\begin{equation}
\nu = \alpha c_s h = \frac{\alpha k_b T(r)}{\mu \Omega_k}, 
\end{equation}
where we have substituted for the isothermal sound speed $c_s = \sqrt{k_b T/\mu}$ and the hydrostatic scale height of the disk $h $ $\simeq c_s/\Omega_k$ (assuming it is vertically isothermal at all $r$) in the second equality. Here, $k_b$ is the Boltzmann constant ($\simeq 1.38 \times 10^{-23}$ J/K) , $\mu $ the mean molecular mass of disk gas ($\simeq 2.4$ proton masses), $T$ the disk midplane temperature, and $\Omega = \sqrt{GM_J/r^{3}}$ the Keplerian angular velocity, $G \simeq 1.67 \times 10^{-11}$ m$^{3}$/kg$\cdot$s$^{2}$ being the gravitational constant. Returning to \textbf{Eq. 2}, $R_H$ represents Jupiter's Hill radius, which roughly defines the disk outer edge. It is given by $a_J(M_J/3M_{\odot})^{1/3}$, where $a_J\simeq 5.2$ AU and $M_{\odot}\simeq 2\times 10^{30}$ kg are Jupiter's semi-major axis and the solar mass, respectively. Towards (proto-)Jupiter, the disk is truncated at the magnetospheric cavity $R_T$, which we take to be $\sim 5$$R_J$. Note that in a viscous decretion disk, the quantity $\nu\Sigma\sqrt{r} \sim \dot{M}$ must decay with $r$, since the radial velocity of gas $v_r$ is directed towards $-d (\nu\Sigma\sqrt{r})dr$. \\ 
\indent As is evident, $\Sigma(r)$ depends on the specification of $T(r)$. Here, we assume an optically thin disk heated solely by viscous shear, for which we have
\begin{equation}
T(r) = \left[\frac{3\dot{M}\Omega_k^2}{16 \pi \sigma_{sb}}\left(\sqrt{\frac{R_H}{r}}-1\right)\right]^{1/4},
\end{equation}
where $\sigma_{sb} \simeq 5.67\times 10^{-8}$ W/m$^{2}\cdot$ K$^4$ is the Stefan-Boltzmann constant. Notably, with this prescription the disk aspect ratio $h/r$ ($\sim \sqrt{T}$; on which the characteristic Type-I damping timescale depends strongly, see \textbf{Section 2.3}) is independent of $\alpha$. \\
\indent An optically thin disk is physically motivated by consideration of the mechanism with which it sources its gas and dust. As noted above, the vertical influx into the circum-Jovian disk (assumed to be confined within $R_T$) is sourced from approximately one scale height above the circumsolar disk midplane, denoted $H$ to avoid confusion with $h$. It is well established that, owing to a balance between turbulent diffusion and gravitational settling, dust particles of a given size settle toward the midplane and establish a sub-disk of scale height $H_d < H$ \textcolor{blue}{(Dubrulle et al., 1995)}. In accord with intuition, the largest particles (\textit{i.e.,} cm-scale and above) in the dust size distribution settle most readily, and are characterized by the lowest $H_d$. Only the smallest (\textit{i.e.,} micron- to 1-mm-scale) particles, constituting a meager fraction of the solid budget, are dispersed to the upper disk layers (\textit{i.e.,} $H_d\sim H$). Thus, gas subsumed into the circum-Jovian disk is expected to be dust-poor \textcolor{blue}{(Tanigawa et al., 2012; Shibaike et al., 2019; Batygin \& Morbidelli, 2020)}, characterized by a metallicity (\textit{i.e.,} dust-to-gas ratio) far smaller than that of the circumsolar disk $\mathbb{Z} \sim 1\%$. \\
\indent For clarity, consider two gas parcels, one at the circumsolar disk midplane, and the other at height $H$ above it. The former contains $\mathbb{Z} \sim1\%$ dust by mass, of which $\sim1\%$ may be held in micron-scale particles to which most of the disk opacity is attributed. This parcel possesses a ``micron-scale" metallicity $\mathbb{Z}_{\mu} \sim 0.01\%$. The latter parcel is also characterized by  $\mathbb{Z}_{\mu} \sim 0.01\%$, since the density of both gas and micron-scale dust are reduced by $\sqrt{e}$ ($H_d\sim H$). However, here $\mathbb{Z} \sim \mathbb{Z}_{\mu}$, as settling of particles for which $H_d<H$ has largely stripped the parcel of solids. \\
\indent Within the circum-Jovian disk, dust coagulation may further diminish $\mathbb{Z}_{\mu}$ \textcolor{blue}{(Mosqueira \& Estrada, 2003; Dullemond \& Dominik, 2005; Batygin \& Morbidelli, 2020)}. If sufficiently rapid, dust accumulation would occur at and near the centrifugal radius (\textit{i.e.,} the infall point), such that only a small fraction of micron-scale particles are advected outward \textcolor{blue}{(Lubow \& Martin, 2013)}. Moreover, our model is envisioned to operate in the late stages of the circum-Jovian disk, when the moons have accreted enough mass to undergo Type-I migration. At this stage, infall (and thus the decretion rate $\dot{M}$) may have waned substantially from dissipation of the circumsolar disk, such that the circum-Jovian disk, and more specifically its optical depth $\tau$ , is low. \\
\indent Adopting an optically thick disk introduces a factor of $\sim (3\tau/4)^{1/4}$ to $T(r)$ in \textbf{Eq. 4} (Armitage, 2020), where $\tau \sim\mathbb{Z}_{\mu} \Sigma k_d/2$ \textcolor{blue}{(Bitsch et al., 2014)} and the dust opacity $k_d\sim 30$ m$^2$/kg. With $\mathbb{Z}_{\mu}\sim 10^{-4}$, it is clear that for $\Sigma \sim$ a few times $10^{4}$ kg/m$^2$ (applicable to the vicinity of the ice-line in our model), we have $\tau\sim 5$. Thus, neglecting the impact of opacity amounts to underestimating $T$ by merely factor of order unity. While we proceed with the optically thin assumption, we recognize the uncertainties that permit it, and discuss the significant changes its relaxation can have on the details of our work in \textbf{Section 5.2}. At this stage, note that while $\tau$ appears to be a weak control on our disk model and thus simulation results (entering as it does into the expression for $T(r)$ to the $1/4$ power) the Type-I eccentricity damping timescale (see \textbf{Section 2.3} below) on which the stability of the system strongly depends scales as $\sim T^2$. Thus, an increase in a factor of a few in $T$ (due to say, $\mathbb{Z}_{\mu}$ being closer to $10^{-3}$ than $10^{-4}$), leads to a commensurate decrease in the efficiency of eccentricity damping.\\

\subsection{Implementing a Pressure Bump}
\indent A steady-state disk, as described above, is one wherein $\dot{M} \sim \Sigma/\nu$ is invariant with $r$ (\textit{i.e.,} expressing negligible buildup/loss of mass in any disk annulus). As such, the functional form of $\Sigma(r)$, beyond a decay with $r$ facilitated by $\Omega_k$, is set wholly by that of $\alpha(r)$. More specifically, any local decrease in $\alpha$ must be counteracted with an increase in $\Sigma$, so as to maintain a constant $\dot{M}$ across the disk. Since the disk midplane pressure $P = \rho_g c_s^2$, where the gas density $\rho_g = \Sigma/(\sqrt{2\pi}h)$, a bump in $P$ is equivalent to one in $\Sigma$, and can be implemented as a dip in $\alpha$. \\
\indent In our model, the dip in $\alpha(r)$ takes the form of an inverted Gaussian centered on $r_0$, the radial location of the $P$ bump. Its minimum is denoted $\alpha_0$, and moving away from $r_0$, $\alpha$ rises and plateaus at a constant value $\alpha_c \sim 10^{-3}$. With these specifications, $\alpha(r)$ is given by 
\begin{equation}
\alpha(r) = \alpha_0 10^{\beta(r)},
\end{equation}
where
\begin{equation}
\beta(r) =\log_{10}\left(\frac{\alpha_0}{\alpha_c}\right)e^{(r-r_0)^2/2w^2} + \log_{10}\left(\frac{\alpha_c}{\alpha_0}\right).
\end{equation}
The width of the Gaussian $w$ reflects that of the $P$ bump, and must be $\gtrsim h_0$ (the hydrostatic scale height at $r_0$) for the bump to be stable against Rossby wave instabilities \textcolor{blue}{(Li et al., 2000; Dullemond et al., 2018)}, and thus a long-lived disk feature. The structure of the $\Sigma/P$ bump controls its ability to halt the Type-I migration of a moon close to its peak, and is set by both $w/h_0$, and  ratio $\alpha_c/\alpha_0$ (specifying its ``height"), henceforth denoted as:
\begin{equation}
R_{\alpha} = \alpha_c/\alpha_0.
\end{equation}
We treat $R_\alpha$ and $w/h_0$ as free parameters, with fiducial values of 2.5 and 1.25 (see \textbf{Section 4.1}), and 2 and 1 (see \textbf{Section 4.3)}, respectively. In exploring their impact on the dynamical evolution of the Galilean moons (see \textbf{Section 4.2}), the former is allowed to range from 1.5 to 5, and the latter from 1 to 2.5.\\
\indent Given a specification of $R_\alpha$ and $w/h_0$, an aspect ratio for the bump can be calculated. To do so requires a translation of $R_\alpha$ to a length scale representative of bump height. The scale height $h_0$ alone is inadequate\textemdash as mentioned above, the assumption of an optically thin disk renders $h$ oblivious to changes in $\alpha$, and thus the $P$ bump. To proceed, note that while the height at which the midplane $P$ falls by $\simeq e^{-1/2}$ does not change with $\alpha$,\textit{ $P$ itself does}. Stated differently, while $h_0$ is constant with $R_\alpha$, $P(z = h_0)$ is not. The bump simply shifts all $P$ contours away from the midplane. That said, we define the height of the bump $\Delta h$ as the difference between $h_0$ and the height corresponding to $P(z=h_0)$ in the absence of the bump, which lies slightly above $h_0$.  The expression for $\Delta h$ is (see \textbf{Appendix})
\begin{equation}
\Delta h = h_0\left(\left[2 ln(R_\alpha\sqrt{e})\right]^{1/2} - 1\right).
\end{equation}
Accordingly, the aspect ratio is defined as $\Delta h/w$.

\begin{figure*} 
\centering
\scalebox{0.6}{\includegraphics{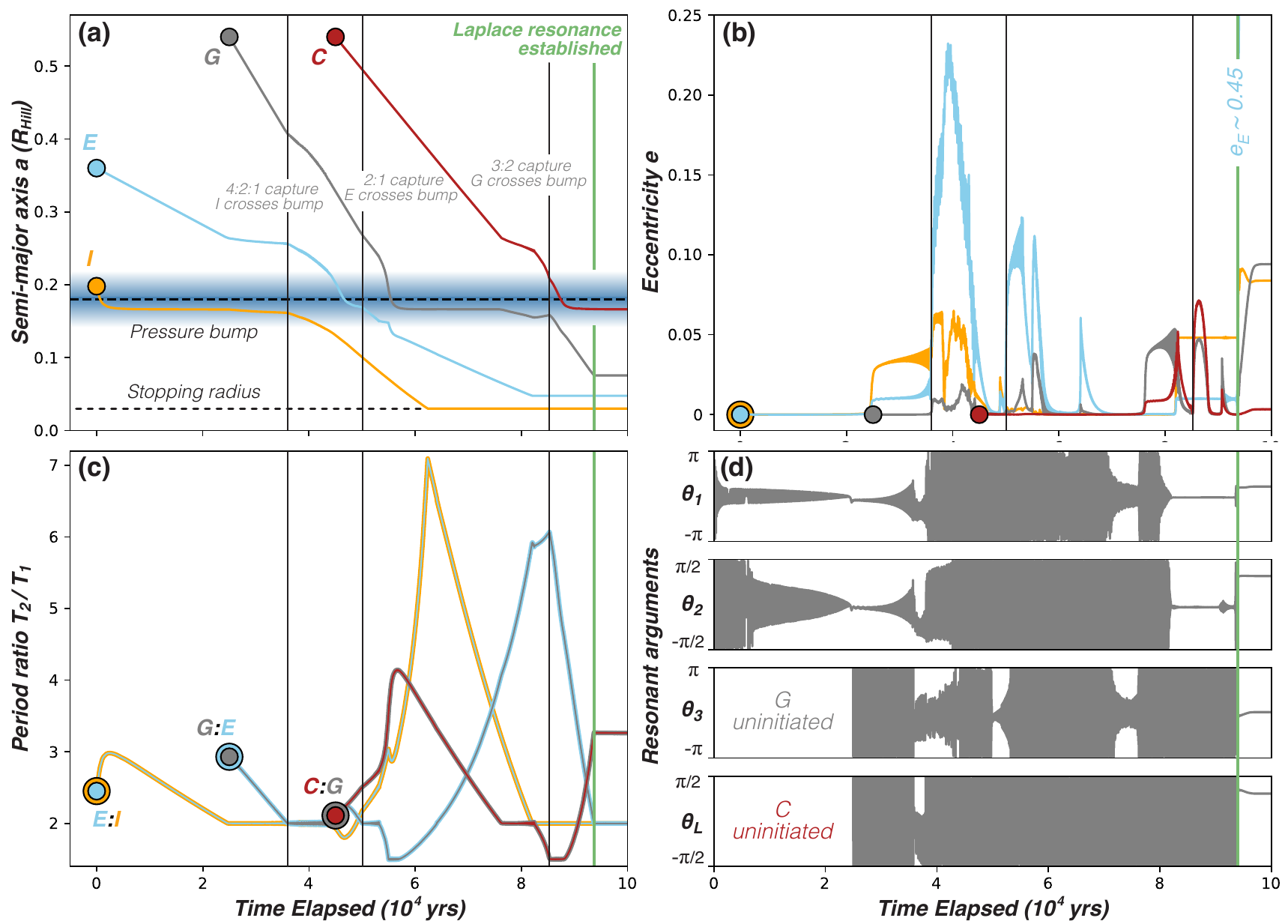}}
\caption{\textbf{Simulation results for $R_{stop} = 0.03 R_{Hill} > R_T = 5 R_J$ and $r_0 = 0.18 R_{Hill}$, assuming fiducial $R_{\alpha} = 2.5$ and $w=1.25 h_0$. }Panels indicate the \textbf{(a)} semi-major axes and \textbf{(b)} eccentricities of the moons, \textbf{(c)} their outer-inner period ratios, and \textbf{(d) }resonant arguments between Io, Europa, and Ganymede. Key resonant captures are denoted by vertical lines. See \textbf{Section 4.1} for discussion.}
\label{fig:Figure 2}
\end{figure*}

\subsection{Type-I Damping}
\indent Comprehensive investigations of planet-disk interactions require resource-intensive hydrodynamic simulations \textcolor{blue}{(\textit{e.g.,} Cresswell \& Nelson, 2008; Bitsch \& Kley, 2010; Pichierri et al., 2023, 2024b)}. When the focus of study lies not in the detailed nature of such interactions, but instead their phenomenology (\textit{e.g.,} how they sculpt the architecture of a planetary system), as is the case here, a more viable avenue is to rely on N-body simulations wherein fictitious forces mimicking the dynamical impact of disk material are implemented. Having constructed the steady-state surface density profile $\Sigma(r)$, we now turn to the Type-I forces it underpins\textemdash recall that satellite migration and $e$-damping are driven by torques exerted on the satellite by the local, perturbed gas. In our simulations, these forces are introduced as operators through REBOUNDx (see\textbf{ Section 3}; \textcolor{blue}{Tamayo et al., 2020}). Here, we outline the key equations used to compute $\dot{a}$ and $\dot{e}$ for each moon. \\
\indent The evolution of $a$ and $e$ under the action of Type-I forces can be expressed in terms of their respective timescales (\textit{i.e.,} $\tau_a$ and $\tau_e$) as
\begin{equation}
\frac{\dot{a}}{a} = -\frac{1}{\tau_a}; \frac{\dot{e}}{e} = -\frac{1}{\tau_e}.
\end{equation}
In terms of the evolution timescale $\tau_m = \dot{\mathcal{L}}/\mathcal{L}$, where $\mathcal{L}$ is the angular momentum, $\tau_a$ takes the form 
\begin{equation}
\tau_a = \left(\frac{2}{\tau_m} + \frac{2e^2}{(1-e^2)\tau_e}\right)^{-1}.
\end{equation}
(This relationship is derived in the \textbf{Appendix}) Formulae for $\tau_m$ and $\tau_e$ are expressed in terms of the the characteristic Type-I damping timescale $\tau_{wave}$, given by \textcolor{blue}{(Tanaka et al., 2002; Tanaka \& Ward, 2004)}
\begin{equation}
\tau_{wave} = \frac{M_J^2}{m_X \Sigma a_X^2 \Omega_{k}}\left( \frac{h}{r}\right)^4 .
\end{equation}
Here, $m_X$ and $a_X$ represent the mass and semi-major axis of moon \textit{X} (\textit{I}, \textit{E}, \textit{G}, or \textit{C}). Notably, $\tau_{wave}$ is shorter for larger $m_X$ and $\Sigma$, with consequences for resonant capture (see \textbf{Section 4}). The timescale $\tau_m$ is given by
\begin{equation}
\tau_m = \frac{\tau_{wave}}{(2.7 + 1.1\gamma)} \left(\frac{h}{r}\right)^{-2} P_e,
\end{equation}
where $\gamma$ is the local power law index of $\Sigma(r)\sim r^{-\gamma}$ (\textit{i.e.,} its slope in log-log space evaluated at the instantaneous position of the body considered). Note that $\gamma$ controls the direction of migration (\textit{i.e.,} the sign of $\tau_m$), and underlies the function of the pressure bump as a ``trap" (on the interior side of the bump, $\gamma$ is negative, leading to outward migration). Physically, this reflects the enhancement of the positive corotation torque, which (down-slope of the bump) overcomes the negative Lindblad torque (see \textbf{Section 5.2.2}). The quantity $P_e$ is obtained from fitting 3D hydrodynamic simulations of protoplanet disk-driven migration \textcolor{blue}{(Cresswell \& Nelson, 2008)}, and takes the form
\begin{equation}
P_e = \frac{1 + \left(\frac{e_X r}{2.25 h}\right)^{1.2} + \left(\frac{e_X r}{2.84 h}\right)^6}{1- \left(\frac{e_X r}{2.02 h}\right)^4 },
\end{equation}
where $e_X$ is the eccentricity of moon \textit{X}. All $r$-dependent disk parameters (\textit{i.e.,} $\Sigma$, $\Omega_k$, $h$) are, like $\gamma$, evaluated at the instantaneous position of the body. Like $\tau_m$, the expression for $\tau_e$ is also refined by best fits to the said hydrodynamic simulations, yielding 
\begin{equation}
\tau_e = \frac{\tau_{wave}}{0.780}\left[1 - 0.14\left(\frac{e_X r}{h}\right)^2 + 0.06\left(\frac{e_X r}{h}\right)^3 \right].
\end{equation}
\indent With these prescriptions, $\dot{a}_X$ and $\dot{e}_X$ are computed at each time step $dt$ (see \textbf{Section 3}), and the variations $\dot{a}dt$ and $\dot{e}dt$ are superposed with those resulting from gravitational interactions (\textit{e.g.,} resonant ``pumping" of eccentricities). 

\section{Simulation Setup}
\indent Our N-body simulations are performed using the WHFAST symplectic Wisdom-Holman integrator in the REBOUND package \textcolor{blue}{(Rein \& Liu, 2012, Rein \& Tamayo, 2015)}, with a time step $dt$ set to 5\% the orbital period of \textit{I}, the innermost moon. Here, we describe their relevant parameters, namely the initial conditions of each moon, the position of the pressure bump, and the distance from (proto-)Jupiter at which we halt migration. The latter, denoted $R_{stop}$, is implemented simply by asserting that the direction of \textit{I}'s migration past that point is reversed (\textit{i.e.,} flipping the sign of $\tau_m$, and thus $\dot{a}$, as calculated with \textbf{Eqs. 9 \& 10} above). \\
\indent We first performed simulations in which $R_{stop}$ was set to $0.03 R_{Hill}$, a factor of $\simeq 4.5$ larger than $R_{trunc} \simeq 0.0067 R_{Hill}$, where migration is expected to have ceased in reality. \hl{This choice for $R_{stop}$ is not physically motivated, but serves to} reduce runtime, as these illustrative simulations make up the $R_{\alpha}-w$ space (see \textbf{Section 2.2)} exploration in \textbf{Section 4.2}. \hl{The only constraint on $R_{stop}$ is that it needs to be sufficiently far within the pressure bump such that resonant capture thereat (\textit{i.e.,} the establishment of the Laplace resonance) does not interfere with dynamics at the bump.} Here, we positioned the pressure bump ($r_0$) at $0.18 R_{Hill}$, far beyond the ice-line to which it may owe its origin. The bump need not be associated with the ice-line, however, and could have emerged at any ``dead zone" where turbulence is reduced (see \textbf{Section 5.1}). The total duration of each simulation is $100$ kyr, and the time at which each moon is initialized $t_X$ is as follows: $t_I =  t_E=0$, $t_G = 25$ kyr, and $t_C = 45$ kyr. The initial semi-major axes of each moon $a_{i,X}$ is defined relative to $r_0$, and given by $a_{i,I}/r_0 = 1.1$, $a_{i,E}/r_0 = 2$, $a_{i,G}/r_0 = a_{i,C}/r_0 = 3$.  As capture into first-order mean motion resonances does not depend on inclination $i$ (\textit{i.e.,} it is absent in the linear expansion of the disturbing function; \textcolor{blue}{\textit{e.g.,} Batygin, 2015}), the moons are initiated with $i=0$, and the system remains planar across the simulation. With eccentricities $e_X$ set to 0 at $t_X$, all other orbital parameters (\textit{i.e.,} $\varpi_X$) are left undefined. The fiducial case discussed in \textbf{Section 4.1} is characterized by $R_{\alpha} =2.5$ and $w=1.25 h_0$. \\
\indent  Following the simulations just described, we performed one other with $R_{stop} = R_{trunc}$ and $r_0$ just beyond (\textit{i.e.,} a factor of 1.25) the ice-line (see \textbf{Section 4.3)}. This simulation, more representative of how we envision the formation of the Laplace resonance, involves a pressure bump with $R_{\alpha} =2$ and $w= h_0$. As discussed further below, given the smaller $r_0$ (and thus the overall higher $\Sigma$ at the bump), the bump must be smaller to accommodate a stable replacement of \textit{E} by \textit{G} at its peak. In other words, having captured \textit{E} in resonance and proceeding to shepherd it across the bump via migration in lockstep, the torque (dependent on $\Sigma$; see \textbf{Eq. 11}) on \textit{G} must not be so large as to destabilize the resonance. The total duration of this simulation is $60$ kyr, and the moons are initialized at $t_I =  t_E=0$, $t_G = 15$ kyr, and $t_C = 25$ kyr. Respective $a_{i,X}/r_0$ remain the same, and as in the ``reduced runtime" simulations above,  $e_X$ and $i_X$ were set to 0 at $t_X$. \\
\indent \hl{The satellite introduction times $t_{E,G,C}$, in addition to disk structure, control the timescale on which our proposed scenario unfolds. Importantly, this timescale, ultimately set by $t_C$, cannot exceed the expected lifetime of the circum-Jovian disk. Broad bounds on the start and end of the disk can be gleaned from cosmochemistry. Regarding the latter, paleomagnetic investigation of meteorites suggests the circumsolar disk, the feedstock for Jupiter's runway gas accretion and the circum-Jovian disk, dissipated at $\sim 4$ Myr from CAI formation (Wang et al., 2017; Borlina et al., 2022). Regarding the former, analyses of nucleosynthetic isotope anomalies in SS materials spanning a wide range (\textit{i.e.,} a few Myr) of inferred accretion ages indicate an early (\textit{i.e.,} within $\sim 1$ Myr post-CAIs) separation of inner and outer SS solid reservoirs, widely thought to have been facilitated by Jupiter's formation (Kruijer et al., 2017; Kleine et al., 2020; Yap \& Tissot, 2023; Tissot et al., 2025). Thus, a circum-Jovian disk lifetime on the order of $\sim 1$ Myr is not implausible. As shown below, our choice of $t_C$ in the tens of kyr leads to the establishment of the Laplace resonance within $\sim 0.1$ Myr, concordant with the said cosmochemical constraints.}

\section{Results}
\subsection{A Fiducial Case}
\indent Before exploring how the structure of the pressure bump determines the final architecture of the Jovian system (see \textbf{Section 4.2}), it is worthwhile to consider in detail a fiducial case for which (i) the 4:2:1 Laplace resonance is established, and (ii) \textit{C} is trapped at the bump. That is, an example of a $R_{\alpha}$ and $w$ pair leading to the desired outcome (for the given $r_0$ and disk model). In \textbf{Fig. 2}, we provide the results from such a case, with $R_{\alpha} = 2.5$ and $w = 1.25 h_0$. The four panels depict the \textbf{(2a)} semi-major axes and \textbf{(2b)} eccentricities of the moons, their \textbf{(2c)} outer-inner period ratios, and \textbf{(2d)} the resonant arguments $\theta_{1-3,L}$ (see \textbf{Section 1}), across the simulation. \\
\indent Soon after the start of the simulation, \textit{I} is trapped at the pressure bump. At $\sim 25$ kyr (when \textit{G} is introduced), \textit{E} captures \textit{I} in a 2:1 resonance, ``pushing" it across the bump in lockstep, albeit weakly given its small mass (see \textbf{Section 2.3}). At $\sim 35$ kyr, \textit{G} captures \textit{E} into a 2:1 resonance, increasing the total torque on the three-body resonant chain and rapidly moving \textit{I} past the bump. The steep, and thus rapid, climb of \textit{E} up the bump at $\sim 45$ kyr (when \textit{C} is introduced) breaks the resonant chain, and leaves \textit{E} trapped at its peak while \textit{I} proceeds to migrate inward toward $R_{stop}$. In $\sim 5$ kyr, \textit{G} recaptures \textit{E} into a 2:1 resonance, and pushes it across the bump, leaving itself trapped. At $\sim 75$ kyr, \textit{C} captures \textit{G} into a 2:1 resonance. As \textit{G} is ushered down the bump, the torque driving it outward (\textit{i.e.,} back toward the peak) increases (due to the steepening profile), and eventually breaks the resonance. At approximately the same time, \textit{E} enters the 2:1 resonance with \textit{I} at $R_{stop}$. At $\sim 85$ kyr, \textit{C} recaptures \textit{G} into the (stronger) 3:2 resonance, and successfully displaces \textit{G} short of $\sim 90$ kyr. Finally, while \textit{C} remains at the bump, \textit{G} enters the 2:1 resonance with \textit{E}, establishing the 4:2:1 Laplace resonance before 100 kyr.\\
\indent With each resonant capture described above, eccentricities are pumped and stabilized by Type-I $e$-damping. At the formation of the Laplace resonance, $e_I\sim 0.08$, $e_E\sim 0.45$, and $e_G\sim 0.09$. These eccentricities are sufficiently high as to induce asymmetric librations of resonant arguments away from $0$ and $\pi$ \textcolor{blue}{(Peale \& Lee, 2002; Batygin \& Morbidelli, 2020)}, as observed in \textbf{Fig. 2d}. Following disk dissipation, tidal dissipation assumes the role of $e$-damping, and without Type-I migration ``forcing" the moons into deeper resonance, eccentricities are rapidly damped (to $\ll0.01$). This leads to the Laplace resonance observed today. 

\begin{figure} 
\scalebox{0.6}{\includegraphics{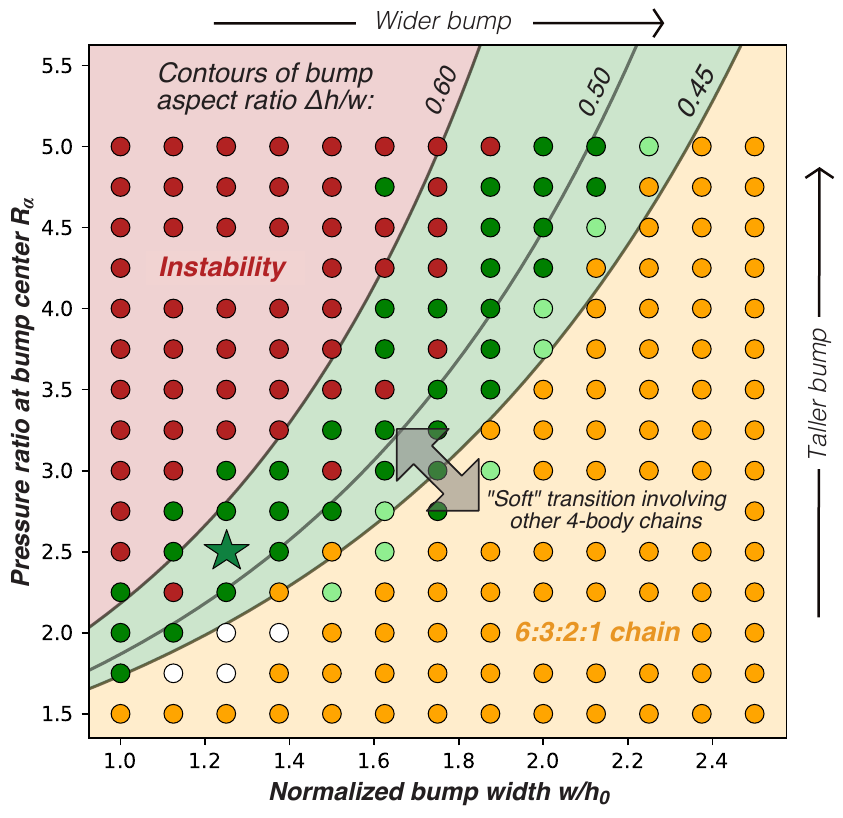}}
\caption{\textbf{Final simulation outcomes from exploration of $R_{\alpha}-w$ parameter space.} Three main regimes are discerned: (i; red) for thin and tall disks to the top left, the bump is too ``stiff" a trap, ultimately leading to a dynamical instability; (ii; yellow) for short and wide disks to the bottom right, the bump fails to trap the moons, leading to the formation of a 6:3:2:1 resonant chain; (iii; green) between (i) and (ii) lies a ``goldilocks" zone wherein the bump can both function as a trap, and allow the moons to be ``pushed" across it by migration in lockstep following resonant capture. This zone corresponds roughly to bump aspect ratios $\Delta h$ between 0.45 and 0.6. Along its former boundary are simulations wherein alternatives to the resonant chain in (ii) are realized (light green: 8:4:2:1; and white: \textit{e.g.,} 12:6:3:2). The green star corresponds to the fiducial case discussed in \textbf{Section 4.1}. See \textbf{Section 4.2} for discussion.}
\label{fig:Figure 3}
\end{figure}

\subsection{Impact of Bump Structure}
\indent We have demonstrated, for a fiducial pressure bump ``height" $R_{\alpha}$ and width $w$, that \textit{I}, \textit{E}, and \textit{G} can be sequentially trapped at the bump, and stably ``pushed" across it via resonant capture and subsequent lockstep migration. Having displaced \textit{G} from the bump, \textit{C} remains trapped while the Laplace resonance is established. We now explore how variations in $R_{\alpha}$ and $w$ modify the simulation outcome. In doing so, we map out three main regimes in parameter space, and show that they can be understood as reflecting a single governing parameter\textemdash the bump aspect ratio $\Delta h$ (see \textbf{Section 2.2} \& \textbf{Eq. 8}). \\
\indent Allowing $R_{\alpha}$ to range between 1.5 and 5 in increments of 0.25, and $w/h_0$ between 1 and 2.5 in increments of 0.125, a total of 195 simulations were performed (including the fiducial case from \textbf{Section 4.1}; \textbf{Fig. 2}). The final results of these simulations are summarized in \textbf{Fig. 3}. The said regimes are easily discerned, and can be intuitively understood. As the bump gets wider and/or shorter (\textit{i.e.,} as we move to the bottom right of parameter space), it loses its function as a migration trap. Accordingly, all four moons make it past the bump, establishing a 6:3:2:1 resonant chain (see \textbf{Fig. 5, S1} for example with $R_{\alpha}=2.5$ and $w = 2h_0$). The 3:2 resonance between \textit{E} and \textit{G} reflects the large mass and thus rapid migration of the latter, violating the adiabatic criterion for 2:1 resonant capture (\textit{i.e.,} ``overshooting" it; \textcolor{blue}{Batygin, 2015}). Conversely, as the bump gets thinner and/or taller (\textit{i.e.,} as we move to the top left of parameter space), it becomes too effective, or ``stiff," as a trap, such that the moons cannot be ``pushed" across. Here, the moons pile up in a resonant chain at the bump, and Type-I migration pumps eccentricities until instability sets in, resulting in either (i) a collision (see \textbf{Fig. 6, S2} for example with $R_{\alpha}=4.5$ and $w = 1.25h_0$), (ii) ejection from the system, or (iii) orbital exchanges (between \textit{E} and \textit{G}) \textcolor{blue}{(Cresswell \& Nelson, 2008)}. In between the two regimes just described lies a ``goldilocks" zone, wherein the bump is ``semi-permeable" and thus conducive to both trapping and migration across it following resonant capture. The general sequence of events in these simulations is consistent with that described in \textbf{Section 4.1}. \\
\indent Overlaid on \textbf{Fig. 3} are contours of $\Delta h/w$. As is evident, these contours nicely bound the trend of the ``goldilocks" zone, demarcating boundaries between the three regimes discussed and indicating that (for our choice of $r_0$, see \textbf{Section 3}) the desired outcome is obtained for $\Delta h/w$ roughly between $0.45$ and $0.6$. Qualitatively, if the bump is widened (\textit{i.e.,} $w$ increases), a complementary increase in its``height" (\textit{i.e.,} $R_{\alpha}$) is needed for the bump to serve its intended purpose. While easily understood, we recognize that $\Delta h/w$ is simply a proxy for $\gamma$ (the local $\Sigma$ power law index; see \textbf{Section 2.3}) at the steepest point along the interior side of the pressure bump, which is the true factor controlling its function. \\
\indent Regime boundaries (where $\Delta h/w \sim 0.6$ and $0.45$) are clearly ``soft" considering simulation outcomes in their vicinity. Regarding the former, it is clear that simulations for which $\Delta h/w < 0.6$ can still result in instability. Close examination of $R_{\alpha}-w$ pairs corresponding to these simulations is hardly a meaningful exercise, as they reflect a confluence of assumed model parameters and initial conditions that yield the unfavorable outcome. While useful to first-order, $\Delta h/w$ alone does not determine the final architecture of the system. Along the latter boundary are simulations resulting in four-body resonances distinct from the 6:3:2:1 chain described above. For $ w\gtrsim 1.5 h_0$ (or $R_{\alpha}\gtrsim 2$), these simulations (light green in \textbf{Fig. 3}) yield, or are evolving toward, an 8:4:2:1 resonance by $100$ kyr (see \textbf{Fig. 7, S3} for example with $R_{\alpha}=3.75$ and $w = 2h_0$). Notably, in these simulations the eccentricities of all four moons, as well as the arguments $\theta_3$ and $\theta_L$, exhibit large amplitude librations once the four-body resonant chain is established. Extending the simulation runtime (to $200$ kyr; for $R_{\alpha}=3.75$ and $w = 2h_0$), we find that the resonant configuration is stabilized. For $ w\lesssim 1.5 h_0$ (or $R_{\alpha}\lesssim 2$), the said simulations (white in \textbf{Fig. 3}) result in either a 12:6:3:2, 16:8:6:3, or (evolution toward a) 9:6:4:2 chain. Before proceeding, we stress that the particularities of these variations on the four-body resonance are insignificant. The key observable remains that a bump too wide for its height fails to keep \textit{C} at bay (and, more subtly, break any 3:2 resonance between \textit{E} and \textit{G}, after the former is ushered off the bump peak; see \textbf{Section 4.1} above). 

\begin{figure} 
\scalebox{0.55}{\includegraphics{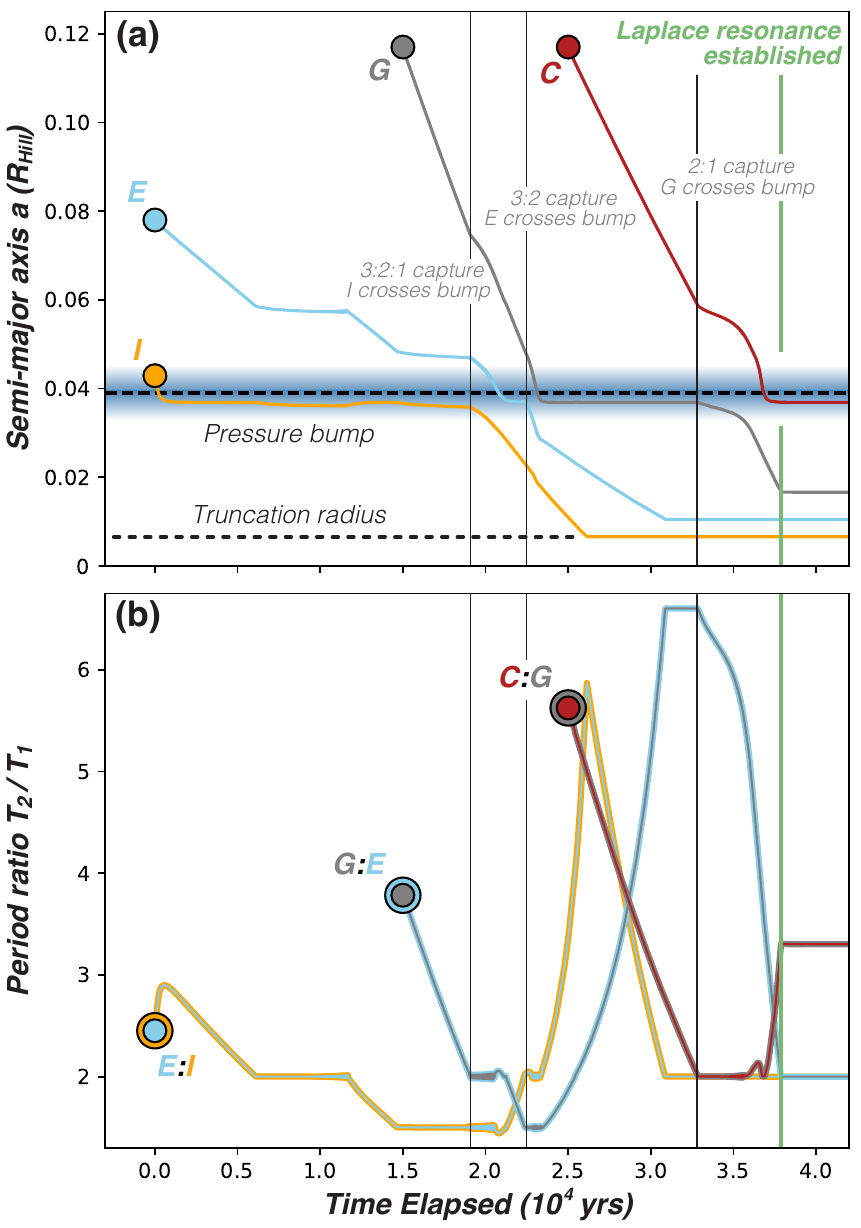}}
\caption{\textbf{Figure 4. Simulation results for $R_{stop} = R_T = 5R_J$ and $r_0$ just beyond (\textit{i.e.,} 1.25 times) the ice-line  ($r_0\sim 0.04 R_{Hill}$), with $R_{\alpha} = 2$ and $w = h_0$. }Panels indicate the \textbf{(a)} semi-major axes of the moons and \textbf{(b)} their outer-inner period ratios. Key resonant captures are denoted by vertical lines. See \textbf{Section 4.3} for discussion.}
\end{figure}

\subsection{A More Representative Run}
\indent As described in \textbf{Section 3}, following our exploration of bump structure, we ran an additional simulation with $R_{stop}$ more aptly coincident with $R_{trunc}$, and $r_0$ positioned slightly beyond the ice-line ($\sim 0.04 R_{Hill}$). This simulation is characterized by $R_{\alpha}=2$ and $w = h_0$, as opposed to $2.5$ and $1.25 h_0$ as assumed in our fiducial case in \textbf{Section 4.1}. Recall that the migration rate is dependent on $\Sigma$ and $\Omega_K$ (\textbf{Eq. 11}). With the pressure bump shifted closer to proto-Jupiter, its underlying surface density $\Sigma \sim r^{-3/4}$ (see \textbf{Eqs. 2, 3, \& 4}) is enhanced by a factor of $\sim (0.04/0.18)^{-3/4} \simeq 3.1$. Moreover, $\Omega_K (r\sim r_0)$ is enhanced by a factor of $\sim (0.04/0.18)^{-3/2}  \simeq 9.5$. Together, this results in a reduction of $\tau_{wave}$ by a factor of $\simeq 30$. Assuming $R_{\alpha}=2.5$ and $w=1.25$, the migration of \textit{G} up the bump (as it displaces \textit{E}) grossly violates the adiabatic criterion, and leads to a collision between the two moons. By shrinking the bump as a whole, stability is restored. Considering our exploration of bump structure in \textbf{Fig. 3}, we infer that for smaller $r_0$, the ``goldilocks" zone is confined to smaller $R_{\alpha}$ and thus $w$: the aspect ratio $\Delta h/w$ must remain approximately the same so as to trap the moons, but the overall bump size ought to decrease to limit the contribution of $\Sigma$ to the shortening of $\tau_{wave}$. This complication is attributed solely to the position of $r_0$ (not $R_{stop}$), which need not be associated with the ice-line. \\
\indent Results from this simulation are shown in \textbf{Fig. 4} (see \textbf{Fig. 8, S4} in the \textbf{Appendix} for plots of $\Sigma(r)$, $T(r)$, $h(r)/r$, and $\tau_{wave}$ corresponding to the simulation). Aside from minor variations in temporary resonances established around the bump, it is apparent that this simulation simply a scaled-down version of that presented in \textbf{Fig. 2} and discussed in \textbf{Section 4.1}. For completeness, we recite the key events here: \textit{I} is trapped at the bump at approximately time zero, \textit{E} captures \textit{I} into a 2:1, then (violating adiabaticity) 3:2 resonance at $\sim 15$ kyr, and \textit{G} captures \textit{E} into a 2:1 resonance at $\sim 18$ kyr. Having displaced \textit{I}, \textit{E} is trapped at the bump until it enters a 3:2 resonance with \textit{G} at $\sim 23$ kyr,. Similarly, having displaced \textit{E}, \textit{G} is trapped at the bump until \textit{C} captures it into a 2:1 resonance at $\sim 33$ kyr. Finally, having displaced \textit{G}, \textit{C} remains trapped at the bump while \textit{I}, \textit{E}, and \textit{G} establish the Laplace resonance interior to it by $\sim 38$ kyr. The eccentricities of all four moons beyond $38$ kyr are essentially the same as those in \textbf{Fig. 2}. As such, the resonant arguments $\theta_{1-3,L}$ librate away from $0$ and $\pi$, as described in \textbf{Section 4.1}. 

\section{Discussion}
\subsection{Pressure Bump Sources}
\indent The origin of pressure bumps in protoplanetary disks is typically ascribed to (i) dust accumulation at sublimation lines, or (ii) so-called ``dead zones" in the disk, wherein turbulence [thought to be sourced by the magnetorotational instability (MRI); \textcolor{blue}{Balbus \& Hawley, 1991}] is quenched. Regarding the former, the ice-line is commonly recognized in semi-analytical and hydrodynamic simulations as a pressure bump source \textcolor{blue}{(\textit{e.g.,} Kretke \& Lin, 2007; Brauer et al., 2008; Bitsch et al., 2015; Charnoz et al., 2021; Müller et al., 2021)}, owing to the buildup of dust, and thus opacity, it induces. Dust buildup likely results in a ``cold finger" effect, involving the (re-)condensation of water vapor diffused across the ice-line following ice sublimation off of inward-drifting refractory/silicate particles \textcolor{blue}{(\textit{e.g.,} Stevenson \& Lunine, 1988; Ros \& Johansen, 2013)}, coupled with the drastic difference between the size (and thus inward drift rates) of silicate and icy dust particles. The latter reflects the higher fragmentation threshold velocity expected for icy particles \textcolor{blue}{(Blum \& Münch, 1993; Güttler et al., 2010; Müller et al., 2021; Batygin \& Morbidelli, 2022; Yap \& Batygin, 2024)}, and results in a so-called ``traffic jam" effect at the ice-line \textcolor{blue}{(Drażkowska \& Alibert, 2017)}. \\
\indent Dead zones of disk turbulence can result from a low ionization fraction (\textit{i.e.,} insufficient coupling between disk gas and magnetic field) in conjunction with non-ideal magnetohydrodynamic (MHD) effects, such as ambipolar diffusion \textcolor{blue}{(Gammie, 1996; Hasegawa \& Pudritz, 2010; Okuzumi \& Hirose, 2011; Yang et al., 2018; Flock et al., 2015; Pinilla et al., 2016, Béthune et al., 2017; Riols \& Lesur, 2018)}. In a dead zone, turbulence $\alpha$ values are typically assumed to range between $10^{-5}$ and $10^{-4}$, an order of magnitude or two less than those characterizing MRI-active disk regions (\textit{i.e.,} $10^{-3}$ to $10^{-2}$). Given the $\alpha$ we adopted for $r<<r_0<<r$ (\textit{i.e.,} away from the bump; $\alpha_c = 10^{-3}$), it appears the values of $R_\alpha$ we explored ($< 6$; see \textbf{Fig. 3}) are not applicable to pressure bumps arising at dead zones, for which we expect $R_\alpha$ to be at least $\sim 10$. The relatively low $R_\alpha$ values we used correspond more closely to what is expected for bumps due to dust accumulation, in accord with our representative run (see \textbf{Section 4.3}) wherein the bump is roughly coincident with the ice-line. \\
\indent While $R_\alpha$ values in our work may not be characteristic of dead zones, two points render this concern trivial. First, our exploration of bump structure in \textbf{Section 4.2} identified the bump aspect ratio $\Delta h/w$ as the key factor determining bump function, not $R_\alpha$ itself. The trend of the ``goldilocks" zone in \textbf{Fig. 3} suggests that a higher $R_\alpha$, such as that consistent with expectations for dead zones, simply needs to be compensated for with a wider bump (\textit{i.e.,} larger $w$). \hl{To maintain $\Delta h/w \sim 0.5$, a value of $R_{\alpha}\sim 10$ implies a bump width of $\sim 2.7 w$, for instance. Nonetheless, $R_{\alpha}$ (and $w$) cannot increase indefinitely for our proposed scenario to be realized, as higher $R_{\alpha}$ leads to higher $\Sigma$ at the bump, in turn promoting more rapid migration (\textit{i.e.,} lower $\tau_{wave}$; see \textbf{Fig. 8, S4b}).} Although $\tau_e$ decreases with $\tau_a$, too fast a migration will lead to \textit{G} ``overshooting" the 3:2 resonance with \textit{E}, leading to either destabilization or capture in a higher order resonance. In the latter case, the separation between the two moons may not be sufficient to allow \textit{E} to clear the bump once \textit{G} reaches its peak. Note that, if the disk (or bump region) is optically thick (see \textbf{Section 5.2} below), higher $\Sigma$ also translates to higher $T$, and thus $h/r$. This in turn leads to a more dramatic drop in $\tau_e \sim (h/r)^{4}$ than $\tau_a \sim \tau_m \sim (h/r)^{2}$ (see \textbf{Section 2.3}). \\
\indent The second point to bear in mind is the possibility that turbulence is not the primary/only angular momentum transport mechanism in the disk. In particular, MHD winds have recently been identified as a compelling alternative for driving protoplanetary disk evolution \textcolor{blue}{(\textit{e.g.,} Blandford \& Payne, 1982; Tabone et al., 2022; Yap \& Batygin, 2024)}. \hl{In this case, angular momentum transport can be described by a \textit{total}, or \textit{effective} $\alpha$, constituting a sum of contributions from turbulence $\alpha_{\nu}$ and disk winds $\alpha_{DW}$ (\textit{i.e.,} $\alpha = \alpha_{\nu} + \alpha_{DW}$). For a given $\alpha$, if the contribution from winds rival or surmount that of turbulence, the formation of a dead zone would constitute a much smaller reduction in angular momentum transport than would be expected if $\alpha$ was solely sourced from turbulence (\textit{i.e.,} if $\alpha = \alpha_{\nu}$). That is, disk winds buffer against the quenching of turbulence, and a dead zone may still be described by $R_\alpha \lesssim 10$.}\\
\indent Alternatives to the two main mechanisms for pressure bump generation discussed exist in the literature, including a hybrid of the two\textemdash  turbulence reduction \textit{due to} dust accumulation \textcolor{blue}{(Dullemond \& Penzlin, 2018)}, which may result from secular gravitational instabilities \textcolor{blue}{(Takahashi \& Inutsuka, 2014)}\textemdash and zonal flows \textcolor{blue}{(\textit{e.g.,} Johansen et al., 2009)}. In their work proposing the growth of the Galilean moons by pebble accretion, \textcolor{blue}{Shibaike et al. (2019)} also considered the possibility of a pressure bump formed just beyond the orbit of \textit{G}, resulting from it reaching its pebble isolation (\textit{i.e.,} disk gap-carving mass).

\subsection{Relaxing Simplifying Assumptions }

\indent Our work primarily serves as an illustrative/proof-of-concept study exploring the potential of a pressure bump in explaining the non-resonant orbit of \textit{C}. As such, we adopted a disk model that, while underpinned by several simplifying assumptions, captures the fundamental aspects of a steady-state viscous disk. Here, we consider the consequences of relaxing the said assumptions, namely that of an optically thin disk (\textbf{Section 5.2.1}) and that of a locally isothermal disk (\textbf{Section 5.2.2}). \hl{In the optically thick and non-isothermal regime, the dependence of Type-I migration on the thermodynamic state of the disk [\textit{i.e.,} $T(r,z)$] cannot be neglected, and would otherwise be accounted for by the inclusion of the local power law index of $T(r,z)$ (\textit{i.e.,} its radial gradient) in \textbf{Eq. 12}, as well as consideration of radiative transfer in the calculation of the hydrostatic scale height $h$, reflecting $T(z)$.}

\subsubsection{An Optically Thin Disk}
\indent Although the gas feeding the circum-Jovian disk is thought to be dust-poor, and dust growth within the disk will reduce the micron-scale metallicity of this gas (see \textbf{Section 2.1}), it is conceivable that a gradual buildup of solids therein (even if held mainly in mm-scale or larger particles) could fuel the production of micron-scale particles via collisional fragmentation, such that opacity is non-negligible in the steady-state disk. Indeed, dust fragmentation is believed to sustain optically thick protoplanetary disks over millions of years \textcolor{blue}{(Williams \& Cieza, 2011)}. This is combated to some extent by satellitesimal formation and growth, which serve to deplete the disk metallicity. Nonetheless, timescales for these processes remain loosely constrained at best.  \\
\indent A more robust hint at an optically thick disk being more realistic is provided by recent smooth particle hydrodynamic simulations of circumplanetary disks, revealing that those characterized by an aspect ratio $h/r$ short of $\sim 0.2$ are efficiently truncated by tidal forces from the host star \textcolor{blue}{(Martin et al., 2023)}. Only for $h/r\gtrsim0.3$ do these simulations reproduce the extensive outflows seen in previous hydrodynamic simulations of circumplanetary disks \textcolor{blue}{(\textit{e.g.,} Tanigawa et al., 2012)}. Such ``puffy" disks depart significantly from our disk model, wherein $h/r = \sqrt{k_b T/\mu v_k^2}$ evaluates to $\lesssim 0.1$ across the disk (see \textbf{Fig. 8, S4b}), given parameter values in \textbf{Section 2.1}. Outflow strength, however, is not solely governed by $h/r$, and is observed to increase with $\alpha$. Note that for $h/r >0.1$, the thin-disk, \hl{vertically isothermal} approximation with which we calculate $h/r$ breaks down, and it is not straightforward to determine how much $T$ (\textit{i.e.,} $\tau$ or $\dot{M}$) ought to increase to yield an increase in $h/r$ by a factor of $\sim3$ \hl{to resist tidal truncation}. \\
\indent As mentioned in \textbf{Section 2.1}, a difference of an order of magnitude in the (largely uncertain) micron-scale metallicity $\mathbb{Z}_\mu$ translates to an increase by a factor of a few in $\tau_e \sim (h/r)^{4} \sim T^2 \sim \sqrt{\mathbb{Z}_\mu}$ (\hl{\textbf{Eq. 11 \& 14};} note that the thin-disk approximation is far less affected, since $h/r \sim \sqrt{T} \sim (\mathbb{Z}_\mu)^{1/8}$). The increase in $\tau_e$ exceeds that of $\tau_a \sim (h/r)^2 \sim (\mathbb{Z}_\mu)^{1/4}$ \hl{(\textbf{Eq. 10})}, such that equilibrium eccentricities in resonance (reflecting a balance between eccentricity damping and ``pumping" from migration deeper into resonance) are higher. Higher eccentricities beget greater instability, implying that for optically thick disks (or larger $\tau$), the size of the ``goldilocks" zone (see \textbf{Section 4.2}) will shrink. We emphasize that our model is envisioned to operate in the late stages of Jupiter's runaway accretion, and in $\tau \sim \mathbb{Z}_\mu \Sigma k_d/2$ \hl{(see \textbf{Section 2.1})}, higher $\mathbb{Z}_\mu$ can be compensated for by lower $\Sigma$ (\textit{i.e.,} a less massive disk overall). Lower $\Sigma$ can be accomplished via either more vigorous turbulence (\textit{i.e., }higher $\alpha$) or a smaller mass decretion/accretion rate $\dot{M}$. In the inner regions circum-Jovian disk ($r< 0.1 R_H$), it is conceivable that Reynolds stresses generated by infall render $\alpha>10^{-3} = \alpha_c$. Regarding $\dot{M}$, the value of $0.1 M_J/Myr$ is assigned \textit{a priori}, and remains a highly uncertain quantity.\\
\indent As a final caveat, we note that whether or not the disk is optically thick or thin in the region of interest may depend on the direction of gas flow therein, as (i) micron-sized particles are tightly coupled to disk gas, and (ii) the steady-state disk metallicity is mediated in part by the interplay of advection, diffusion, and headwind drag \textcolor{blue}{(\textit{e.g.,} Batygin \& Morbidelli, 2020)}. Here, the description of the circum-Jovian disk as decreting beyond the magnetospheric truncation radius reflects our current understanding of circumplanetary disks. Nonetheless, it is conceivable that the centrifugal radius of infall (dependent on the angular momentum budget of gas subsumed into the circum-Jovian disk) was farther out, such that the disk is accretionary in the region of interest. Foreshadowing the discussion below, decretion and accretion disks will likely lead to differing distributions of micron-sized dust, and thus opacity/temperature substructures, capable of shaping the behavior of Type-I migration. 

\subsubsection{A Locally Isothermal Disk}
\indent Related to the issue of opacity is our assumption of a locally isothermal disk\textemdash that is, a disk wherein energy generated from gas compression by the satellite, viscous dissipation, etc. is immediately radiated away, thereby keeping \hl{$T(r) \sim z^0$ and fixed}. As the efficiency of radiation depends strongly on opacity, this assumption is consistent insofar as the disk can be treated as optically thin. \\
\indent The guiding principle in our work has been that Type-I migration is directed toward higher surface density. Encapsulated in this principle is our understanding of the Lindblad torque and the \textit{barotropic}, or vortensity-driven, corotation torque (Goldreich \& Tremaine, 1979; Tanaka et al., 2002). Indeed, the strong dependence of the latter on $(-)\gamma$ (recall $\Sigma(r) \sim r^\gamma$; \hl{see \textbf{Section 2.3 \& Eq. 12}}) is the reason a pressure bump functions as a migration trap. In relaxing the locally isothermal assumption, an additional, thermal contribution to the corotation torque arises (\textit{e.g.,} Paardekooper \& Mellema, 2006, 2008; Paardekooper \& Papaloizou, 2008; Baruteau \& Masset, 2008; Paardekooper et al., 2011), dependent on the radial gradient of entropy/$T(r,z)$. As gas in the corotation region executes horseshoe orbits, it contracts (\textit{i.e.,} is compressed) as it moves toward the inner disk ahead of the satellite, but expands as it moves toward the outer disk behind the satellite. This results in a leading over-density that exerts a positive torque on the satellite, promoting outward migration.  \\
\indent Radiation-hydrodynamic simulations indicate that the thermal corotation torque can halt, or even reverse the inward Type-I migration in the presence of a negative gradient in \hl{$T(r,z)$} when opacities are sufficiently high (\textit{i.e.,} when the disk is locally non-isothermal). This suggests that as opacity increases, (i) wider and/or shorter pressure bumps would be sufficient for satellite trapping (see \textbf{Section 4.2}) since the positive corotation torque has both barotropic and thermal contributions, and (ii) migration toward higher $\Sigma$ is slower, if not completely halted. Moreover, it suggests the possibility for a steep, negative gradient in \hl{$T(r,z)$} as a migration trap\textemdash that is, that pressure bumps are not the sole substructures that can be invoked to explain \textit{C}'s non-resonant orbit. Self-shadowing from a hot, ``puffy" inner rim of the disk, for instance, may engender a dramatic temperature contrast between inner and outer disk regions \textcolor{blue}{(\textit{e.g.,} Dullemond et al., 2001; Schneeberger \& Mousis, 2025)}.\\
\indent While $T(r,z)$ ought to be considered in models of optically thick and non-isothermal disks, the potency of its impact on Type-I migration is not guaranteed. A complication with corotation torques is the possibility of saturation \textcolor{blue}{(\textit{e.g.,} Ogilvie \& Lubow, 2003)}. In essence, gas in the corotation region carries a finite amount of angular momentum that it can exchange with the migrating body. Following this exchange, and in the absence of any (efficient) diffusive process to evacuate the region of angular momentum, the corotation torque is instantaneously zero. In non-isothermal disks, both thermal and viscous diffusion are required to keep the corotation torque unsaturated, and the efficacy of this process in viable models for the circum-Jovian disk (and arguably what is ``viable" in the first place) remains unclear. \\

\subsection{Variations on the Proposed Scenario}
\indent Our simulations show that a pressure bump in the circum-Jovian disk, neither too sharp nor flat (aspect ratio $\Delta h/w \sim 0.5$), can readily exclude \textit{C} from resonance with \textit{I}, \textit{E}, and \textit{G}. In these simulations, the birthplaces of all four moons are assumed to lie beyond the bump, and the three inner moons are sequentially ``pushed" across it by their exterior neighbor. While this is how we envisioned the Laplace resonance came to be, it is worthwhile to note that this picture is conservative with regard to bump structure. Assuming \textit{C} did not form late nor so slow as to render its migration inconsequential (see \textbf{Section 5.2} below), only it had to accrete beyond the bump. Neglecting constraints on composition (\textit{i.e.,} the ice-line position with respect to the bump), \textit{I}, \textit{E}, and \textit{G} could have formed interior to the bump and proceeded to establish the Laplace resonance. The final architecture of the system would be no different from that in our envisioned scenario. If this were the case, there would be no constraint on how sharp the bump can be, barring $w\gtrsim h_0$ for stability (see \textbf{Section 2.2}), since its sole purpose would be to halt the migration of \textit{C}. This amounts to relieving the upper bound on $\Delta h/w$ for the ``goldilocks" zone (see \textbf{Fig. 3}). Alternatives that are equally valid from a solely dynamical standpoint involve the origin of either \textit{G} or \textit{E}$+$\textit{G} outside the bump with \textit{C}. While conceivable, such scenarios incur the drawback of having to invoke multiple satellitesimal formation sites in the disk \textcolor{blue}{(Batygin \& Morbidelli, 2020)}.\\
\indent Considering composition, in particular water-ice content, permits additional insight into the birthplaces of the four moons given a disk model. Gravity data from the Galileo spacecraft reveal that \textit{G} and \textit{C} are characterized by ice-rock ratios close to unity \textcolor{blue}{(Kuskov \& Kronrod, 2001; Sohl et al., 2002)}. Although the source for this abundance of ice remains debated \textcolor{blue}{(Yap \& Stevenson, 2023)}, it is strongly suggested that the two moons originated beyond the ice-line. Similarly, the substantial presence of water-ice in \textit{E} \textcolor{blue}{(Carr et al., 1998; Kivelson et al., 2000)} is most easily explained by icy building blocks, although the delivery of structurally bound water to within the ice-line as phyllosilicates should also be considered \textcolor{blue}{(Ciesla \& Lauretta, 2005)}. As for \textit{I}, it is unclear if the tortured moon accreted ``wet," having since lost its volatiles through volcanism and degassing, or from wholly ``dry" components \textcolor{blue}{(McKinnon, 2023; de Kleer et al., 2024)}. The setup of our simulations, of course, implicitly assumes the former. While the location of the pressure bump is \textit{dynamically} significant, the location of the ice-line is \textit{compositionally} so. In considering variations to the scenario we have painted for assembling the architecture of the Galilean moons, the ice-line sets the minimum jovicentric distance for the birthplaces of \textit{G} (and likely \textit{E}, with the possibility of \textit{I}), but the farther between the ice-line and bump location sets that of \textit{C}. 

\subsection{No need for a late/slow accretion of Callisto}
\indent In the absence of disk substructure (\textit{e.g.,} pressure bumps), a late and/or slow accretion of \textit{C} (\textit{i.e.,} lasting $\sim10^6$ yrs) has been invoked to explain its exclusion from resonance \textcolor{blue}{(Peale \& Lee, 2002; Batygin \& Morbidelli, 2020)}. Underpinning this conjecture is the inferred moment of inertia (MOI) of \textit{C} from three flybys of the moon by the Galileo spacecraft, suggesting it is only partially differentiated, with a core comprising a mixture of rock and ice \textcolor{blue}{(Anderson et al., 1998; 2001)}. To limit interior heating by the short-lived radionuclide $^{26}$Al and accretion, \textcolor{blue}{Barr \& Canup (2008)} suggest \textit{C} must have accreted over a period $\gtrsim 0.5$ Myr and completed its accretion $\gtrsim 4$ Myr after CAI formation (see \textbf{Section 1}), assuming an ambient disk temperature of $\simeq 100$ K.\\
\indent Although widely cited, the partially differentiated interior of \textit{C} remains dubious. Determination of icy satellite MOIs largely rely on the Radau-Darwin approximation (RDA), which relates the MOI of a tidally and rotationally deformed body (of mass $M$ and mean radius $R$) about its spin axis $I$ to its rotation rate $\Omega$ (as encapsulated in its ``centrifugal potential" $q = \Omega^2 R^3/GM$) and shape (as encapsulated in the $J_2$ gravitational coefficient or its flattening $f$) \textcolor{blue}{(Murray \& Dermott, 1999)}. The key assumption in the RDA is that the body is \textit{hydrostatic}. For small and, more pertinently, slow-rotating bodies such as giant planet satellites that (i) lie relatively far from the planet and (ii) rotate synchronously with their orbit (\textit{e.g.,,} \textit{C} and Titan), non-hydrostatic effects can severely undermine the accuracy of the RDA, resulting in erroneous MOIs \textcolor{blue}{(Mueller \& McKinnon, 1988; Gao \& Stevenson, 2013)}. This can be understood dimensionally. Assuming a homogeneous spheroid for simplicity, the distortion from hydrostatic equilibrium is defined as $\epsilon = \sigma/\rho g R$, where $\rho$ and $g$ are the density of the spheroid and gravitational acceleration at its surface, respectively, and $\sigma$ is the deviatoric stress associated with the distortion. The importance of non-hydrostatic effects, then, can be quantified by the dimensionless value $\epsilon/f \sim \epsilon/q \sim \sigma/\rho\Omega^2 R^2 \sim (\Omega R)^{-2}$. Only a $\sim 10\%$ error in the MOI of \textit{C} is necessary for it to be consistent with full differentiation, corresponding to $\sigma\sim 0.1$ bar at its surface or $\sim 1$ bar at its core-mantle boundary \textcolor{blue}{(Gao \& Stevenson, 2013)}. The radio science experiment JUpiter ICy moons Explorer (JUICE) mission \textcolor{blue}{(Grasset et al., 2013)} will ultimately evaluate the magnitude of non-hydrostatic effects and elucidate the interior structure of \textit{C}.\\
\indent A pressure bump alleviates constraints on the formation timing, rate, and location of \textit{C}, and thus circumvents the uncertainty regarding its interior structure. So long as \textit{C} forms beyond the bump (and likely the ice-line; see \textbf{Section 5.1}), it can form earlier, or more rapidly, than inferred based on a supposed need for partial differentiation. 

\section{Concluding Remarks}
\indent In this work, we simulated the Type-I migration of the Galilean moons in the Jovian circumplanetary disk, demonstrating that a pressure bump therein can act as a migration trap for Callisto, preventing its participation in resonance with Io, Europa, and Ganymede. ALMA observations have revealed the ubiquity of concentric dust rings in protoplanetary disks, suggesting pressure bumps are a universal outcome of disk evolution. By invoking a pressure bump, the orbital architecture of the moons is naturally reproduced, and constraints on the timing of Callisto's accretion are eased. Central to unraveling the history of the Jovian system is Callisto's interior structure, of which we eagerly await revelation by JUICE in the coming decade.

\section*{Acknowledgments}
\indent This work was supported by a Caltech Center for Comparative Planetary Evolution (3CPE) grant to the authors, and the David \& Lucile Packard Fellowship to KB. We thank Dave Stevenson for insightful discussions on Callisto's interior structure, and caveats associated with determining icy satellite moments of inertia. 

\section*{Appendix}
\indent \textit{All equations introduced in the main text are boxed.}
\subsection*{Disk model}
\indent The derivation of $\Sigma(r)$ is provided in \textcolor{blue}{Batygin \& Morbidelli (2020)}, and the purpose of its reiteration here is to ensure the paper is self-contained. The foundations of any surface density profile are mass and angular momentum conservation. The former is given by
\begin{equation}
r\frac{\partial \Sigma}{\partial t} + \frac{\partial (r\Sigma v_r)}{\partial r} = S_M ,
\end{equation}
where $v_r$ represents the radial decretionary velocity of gas, and $S_M$ the mass source term. Interested as we are in a steady-state solution, valid so long as the mass decretion rate $\dot{M}$ varies on a longer timescale than that of our simulation, $\partial \Sigma/\partial t \rightarrow 0$.  Moreover, in our region of interest (\textit{i.e.,} between the inner and outer edges of the disk), we may safely assume $S_M = 0$. Implicit to this assumption is that the vertical inflow of gas and dust is confined to within the disk inner edge, set by magnetospheric truncation at $R_T\sim 5R_J$, where $R_J$ is the radius of (proto-) Jupiter. The quantity $r\Sigma v_r$ (\textit{i.e.,} mass flux per radian), and thus $\dot{M} = 2\pi r\Sigma v_r$ (by cylindrical geometry), is thus invariant with $r$. The continuity equation for angular momentum is expressed as 
\begin{equation}
r\frac{\partial (r \Sigma v_k)}{\partial t} + \frac{\partial (r^2\Sigma v_k v_r)}{\partial r} = S_{AM} , 
\end{equation}
where $v_k = \sqrt{G M_J / r^3}$ is the azimuthal Keplerian velocity, $G$ being the gravitational constant and $M_J$ the mass of Jupiter, and $S_{AM}$ is the angular momentum source term. Angular momentum redistribution in the disk, which drives $\dot{M}$, results from differential torques exerted on disk annuli by mutual shear stresses acting on their inner and outer edges. This mechanism is encapsulated in $S_{AM}$, taking the form \textcolor{blue}{(Armitage, 2020)}
\begin{equation}
S_{AM} = \frac{1}{2\pi} \frac{d \Gamma}{dr} = \frac{\partial}{\partial r}\left(r^3 \nu \Sigma \frac{d\Omega_k}{dr} \right).
\end{equation}
Here, $\Gamma(r)$ represents the net torque on an annulus, $\Omega_k = v_k/r$ the Keplerian angular velocity, and $\nu$ the turbulent viscosity. In the Shakura-Sunyaev prescription, the latter is given by $\alpha c_s h$, where the isothermal sound speed $c_s = \sqrt{k_b T/\mu}$  ($k_b$ being the Boltzmann constant, $T $ the disk temperature, and $\mu$ the mean molecular weight of disk gas), and the hydrostatic scale height of the disk $h = c_s/\Omega_k$ assuming it is geometrically thin ($h/r\ll$1) and vertically isothermal. Setting $\partial (r\Sigma v_k)/\partial t \rightarrow 0$ and $r\Sigma v_r = \dot{M}/2\pi$ from mass conservation, and further defining the vertically integrated viscous stress tensor $W = -r\Sigma \nu (d\Omega_k/dr)$, we can rewrite Eq. (16) as 
\begin{equation}
\frac{1}{2\pi}\frac{d}{d r} (\dot{M}r v_k) = -\frac{d}{d r}(r W^2),
\end{equation}
which simplifies to 
\begin{equation}
r^2\frac{dW}{dr} + 2rW + \frac{\dot{M}v_k}{4\pi} = 0.
\end{equation}
Solving for $W(r)$ yields 
\begin{equation}
W(r) = \frac{\dot{M}\Omega_k}{2\pi} + \frac{C}{r^2},
\end{equation}
where $C$ is a constant of integration, which can be obtained by setting $W=0$ at the disk outer edge, taken as Jupiter's Hill radius $R_H = a_J (M_J/3 M_{\odot})^{1/3}$, with $a_J$ as Jupiter's semi-major axis, and $M_{\odot}$ the solar mass. Imposing this boundary condition, we find $C = \dot{M}\sqrt{G M_J R_H}/2\pi$, which, along with $W = (3/2)\nu \Sigma\Omega_k$,  finally yields 
\begin{equation}
\boxed{\Sigma(r) = \frac{\dot{M}}{3\pi\nu}\left( \sqrt{{\frac{R_H}{r}}}-1\right).}
\end{equation}
\indent  The complete expression of $\Sigma(r)$ requires a specification of the disk temperature profile $T(r)$, introduced in the above through $c_s$ in $\nu$. Here, we assume an optically thin, and viscously heated disk. The temperature of such a disk is set via equilibrium between heat generation and radiative loss, expressed as \textcolor{blue}{(Armitage, 2020)}  
\begin{equation}
\sigma T^4 \simeq  F_{visc}, 
\end{equation}
where $\sigma$ is the Stefan-Boltzmann constant. The heating rate per unit area $F_{visc}$ is given by \textcolor{blue}{(Nakamoto \& Nagakawa, 1994)}
\begin{equation}
F_{visc} = \frac{1}{2}\Sigma \nu \left(r \frac{d \Omega_k}{dr}\right)^2 = \frac{3\dot{M}\Omega_k^2}{8\pi}\left(\sqrt{\frac{R_H}{r}}-1\right).
\end{equation}
Upon substitution into Eq. (22) and solving for $T$, we obtain
\begin{equation}
\boxed{T(r) = \left[\frac{3\dot{M}\Omega_k^2}{16 \pi \sigma_{sb}}\left(\sqrt{\frac{R_H}{r}}-1\right)\right]^{1/4}}
\end{equation}
Hence, aside from physical constants, $\Sigma(r)$ is fully specified by a $\dot{M}$ and $\alpha$.
\subsection*{The pressure bump}
\indent As discussed in the main text, for our steady-state (\textit{i.e.,} constant $\dot{M}$) disk, a bump in pressure can be implemented as a dip in $\alpha$. We assume this dip takes the form of a Gaussian, centered on a jovicentric distance $r_{0}$. With the dip minimum denoted $\alpha_0$ and the value of $\alpha$ (far) outside the bump $\alpha_c$, the profile $\alpha(r)$ takes the form 
\begin{equation}
\boxed{\alpha(r) = \alpha_0 10^{\beta(r)},}
\end{equation}
where
\begin{equation}
\boxed{\beta(r) =\log_{10}\left(\frac{\alpha_0}{\alpha_c}\right)e^{(r-r_0)^2/2w^2} + \log_{10}\left(\frac{\alpha_c}{\alpha_0}\right).}
\end{equation}
The ratio $\alpha_c/\alpha_0$ (denoted $R_{\alpha}$) is a proxy for the ``height" of the bump, while $w \gtrsim h_0$ sets its width. To calculate a bump aspect ratio, the translation of $R_{\alpha}$ into a length scale is required. As mentioned in the main text, since $\alpha$ does not enter into the expression for $T$, $h (= k_b T/\mu \Omega_k^2)$ is invariant with respect to $R_{\alpha}$. That is, with or without a pressure bump, $h(r_0) = h_0$ remains constant. Nonetheless, $h$ merely represents the height at which the midplane pressure falls by $\sqrt{e}$. While $h_0$ does not depend on $R_{\alpha}$, the \textit{midplane pressure, and thus the pressure at $h_0$, certainly does.} As such, we can define the physical height of the bump $\Delta h$ as the difference between $h_0$ and the height corresponding to $P(z=h_0)$ in the absence of the bump, $h_0'$. Denoting $P(r_0)$ in the presence (absence) of a bump (\textit{i.e.,} $\alpha$ dip) as $P_0$ ($P_f$), $h_0'$ is defined by $P_0(z=h_0') = P_f(z=h_0)$.\\
\indent Keeping in mind $P = \Sigma c_s^{2}/\sqrt{2\pi}h$, we have [see expression for $\Sigma(r)$ above], 
\begin{equation}
P_0 (z)= P_f (z) R_{\alpha} e^{z^2/2 h_0^2} . 
\end{equation}
Without a bump, the pressure at $h_0$ above the midplane is $P_f /\sqrt{e}$. With a bump, the height at which $P_0 = P_f/\sqrt{e}$, denoted $h_0'$, is thus given by 
\begin{equation}
\sqrt{e}=  R_{\alpha} e^{h_0'^2/2 h_0^2}, 
\end{equation}
which simplifies to
\begin{equation}
h_0' = h_0[2ln(R_{\alpha}\sqrt{e})]^{1/2} . 
\end{equation}
Finally, $\Delta h = h_0'-h_0$ is given by
\begin{equation}
\boxed{\Delta h = h_0\left(\left[2 ln(R_\alpha\sqrt{e})\right]^{1/2} - 1\right),}
\end{equation}
and the bump aspect ratio is defined as $\Delta h/w$. 

\subsection*{Relationship between $\tau_a$, $\tau_e$, and $\tau_m$}
\indent With $\Sigma(r)$ constructed, Type-I $a$- and $e$-damping rates can be computed and implemented for each moon at each time step in our simulations. These rates (\textit{i.e.,} $\dot{a}$ and $\dot{e}$) are expressed in terms of their respective timescales as 
\begin{equation}
\boxed{\frac{\dot{a}}{a} = -\frac{1}{\tau_a};   \frac{\dot{e}}{e} = -\frac{1}{\tau_e}.}
\end{equation}
Damping proceeds through the torque exerted on the moon by the perturbed disk gas, which saps the moon of orbital angular momentum $\mathcal{L}$, given by
\begin{equation}
\mathcal{L} = \mu_{m,X} \sqrt{G(M_J + m_X)a_X(1-e_X^2)}.
\end{equation}
Here, $m_X$, $a_X$, and $e_X$ are the mass, semi-major axis, and eccentricity of moon \textit{X} (\textit{I, E, G, C}), and $\mu_{m,X} = (m_X M_J)/(m_X + M_J)$ the corresponding reduced mass. Similar to $\dot{a}$ and $\dot{e}$ above, the said torque $\Gamma = \dot{\mathcal{L}}$ can be expressed in terms of a migration timescale $\tau_m$ as \textcolor{blue}{(Tanaka et al., 2002; Tanaka \& Ward, 2004)} 
\begin{equation}
\frac{\dot{\mathcal{L}}}{\mathcal{L}} = -\frac{1}{\tau_m}.
\end{equation}
To elucidate the relationship between $\tau_m$, $\tau_a$, and $\tau_e$, we simply evaluate $\dot{\mathcal{L}}$, yielding
\begin{equation}
\Gamma = \dot{\mathcal{L}} = \mathcal{L}\left(-\frac{1}{2\tau_a} + \frac{e^2}{(1-e^2)\tau_e}\right) = -\frac{\mathcal{L}}{\tau_m}.
\end{equation}
Accordingly, 
\begin{equation}
\boxed{\tau_a = \left(\frac{2}{\tau_m} + \frac{2e^2}{(1-e^2)\tau_e}\right)^{-1}.}
\end{equation}
Note that $\tau_a$ represents the characteristic timescale for the evolution in orbital energy $\mathcal{E} = -G(M_J + m_X)\mu_{m,X}/2a_X$, as $\dot{\mathcal{E}} = \mathcal{E}/\tau_a$. Formulae for $\tau_m$ and $\tau_e$ are expressed in terms of the characteristic Type-I damping timescale 
\begin{equation}
\boxed{\tau_{wave} = \frac{M_J^2}{m_X \Sigma a_X^2 \Omega_{k}}\left( \frac{h}{r}\right)^4 ,}
\end{equation}
and informed by fitting 3D hydrodynamic simulations of protoplanet Type-I migration \textcolor{blue}{(Cresswell \& Nelson, 2008)}. They are given in the main text (see \textbf{Section 2.3}).

\section*{References}

Anderson, J. D., Jacobson, R. A., McElrath, T. P., Moore, W. B., Schubert, G., \& Thomas, P. C. (2001). Shape, mean radius, gravity field, and interior structure of Callisto. \textit{Icarus}, \textit{153}(1), 157-161.\\

Anderson, J. D., Schubert, G., Jacobson, R. A., Lau, E. L., Moore, W. B., \& Sjogren, W. L. (1998). Distribution of rock, metals, and ices in Callisto. \textit{Science}, \textit{280}(5369), 1573-1576.\\

Andrews, S. M., Huang, J., Pérez, L. M., Isella, A., Dullemond, C. P., Kurtovic, N. T., ... \& Ricci, L. (2018). The disk substructures at high angular resolution project (DSHARP). I. Motivation, sample, calibration, and overview. \textit{The Astrophysical Journal Letters}, \textit{869}(2), L41.\\

Armitage, P. J. (2020). \textit{Astrophysics of planet formation}. Cambridge University Press.\\

Balbus, S. A., \& Hawley, J. F. (1991). A powerful local shear instability in weakly magnetized disks. I-Linear analysis. II-Nonlinear evolution. \textit{Astrophysical Journal, Part 1 (ISSN 0004-637X), vol. 376, July 20, 1991, p. 214-233.}, \textit{376}, 214-233.\\

Barr, A. C., \& Canup, R. M. (2008). Constraints on gas giant satellite formation from the interior states of partially differentiated satellites. \textit{Icarus}, \textit{198}(1), 163-177.\\

Baruteau, C., \& Masset, F. (2008). On the corotation torque in a radiatively inefficient disk. \textit{The Astrophysical Journal}, \textit{672}(2), 1054.\\

Batygin, K. (2015). Capture of planets into mean-motion resonances and the origins of extrasolar orbital architectures. \textit{Monthly Notices of the Royal Astronomical Society}, \textit{451}(3), 2589-2609.\\

Batygin, K., \& Adams, F. C. (2017). An analytic criterion for turbulent disruption of planetary resonances. \textit{The Astronomical Journal}, \textit{153}(3), 120.\\

Batygin, K., \& Adams, F. C. (2025). Determination of Jupiter’s primordial physical state. \textit{Nature Astronomy}, 1-10.\\

Batygin, K., \& Morbidelli, A. (2020). Formation of giant planet satellites. \textit{The Astrophysical Journal}, \textit{894}(2), 143.\\

Batygin, K., \& Morbidelli, A. (2022). Self-consistent model for dust-gas coupling in protoplanetary disks. \textit{Astronomy \& Astrophysics}, \textit{666}, A19.\\

Batygin, K., \& Morbidelli, A. (2023). Formation of rocky super-earths from a narrow ring of planetesimals. \textit{Nature Astronomy}, \textit{7}(3), 330-338.\\

Béthune, W., Lesur, G., \& Ferreira, J. (2017). Global simulations of protoplanetary disks with net magnetic flux-i. non-ideal mhd case. \textit{Astronomy \& Astrophysics}, \textit{600}, A75.\\

Birnstiel, T. (2024). Dust growth and evolution in protoplanetary disks. \textit{Annual Review of Astronomy and Astrophysics}, \textit{62}.\\

Bitsch, B., Johansen, A., Lambrechts, M., \& Morbidelli, A. (2015). The structure of protoplanetary discs around evolving young stars. \textit{Astronomy \& Astrophysics}, \textit{575}, A28.\\

Bitsch, B., \& Kley, W. (2010). Orbital evolution of eccentric planets in radiative discs. \textit{Astronomy \& Astrophysics}, \textit{523}, A30.\\

Bitsch, B., Morbidelli, A., Lega, E., \& Crida, A. (2014). Stellar irradiated discs and implications on migration of embedded planets-II. Accreting-discs. \textit{Astronomy \& Astrophysics}, \textit{564}, A135.\\

Blandford, R. D., \& Payne, D. G. (1982). Hydromagnetic flows from accretion discs and the production of radio jets. \textit{Monthly Notices of the Royal Astronomical Society}, \textit{199}(4), 883-903.\\

Blum, J., \& Münch, M. (1993). Experimental investigations on aggregate-aggregate collisions in the early solar nebula. \textit{Icarus}, \textit{106}(1), 151-167.\\

Borlina, C. S., Weiss, B. P., Bryson, J. F., \& Armitage, P. J. (2022). Lifetime of the outer solar system nebula from carbonaceous chondrites. \textit{Journal of Geophysical Research: Planets}, \textit{127}(7), e2021JE007139.\\

Brasser, R., \& Mojzsis, S. J. (2020). The partitioning of the inner and outer Solar System by a structured protoplanetary disk. \textit{Nature Astronomy}, \textit{4}(5), 492-499.\\

Brauer, F., Henning, T., \& Dullemond, C. P. (2008). Planetesimal formation near the snow line in MRI-driven turbulent protoplanetary disks. \textit{Astronomy \& Astrophysics}, \textit{487}(1), L1-L4.\\

Burkhardt, C., Dauphas, N., Hans, U., Bourdon, B., \& Kleine, T. (2019). Elemental and isotopic variability in solar system materials by mixing and processing of primordial disk reservoirs. \textit{Geochimica et Cosmochimica Acta}, \textit{261}, 145-170.\\

Canup, R. M., \& Ward, W. R. (2002). Formation of the Galilean satellites: Conditions of accretion. \textit{The Astronomical Journal}, \textit{124}(6), 3404.\\

Carr, M. H., Belton, M. J., Chapman, C. R., Davies, M. E., Geissler, P., Greenberg, R., ... \& Veverka, J. (1998). Evidence for a subsurface ocean on Europa. \textit{Nature}, \textit{391}(6665), 363-365.\\

Cassen, P., Reynolds, R. T., \& Peale, S. J. (1979). Is there liquid water on Europa?. \textit{Geophysical Research Letters}, \textit{6}(9), 731-734.\\

Chambers, J. E. (2009). An analytic model for the evolution of a viscous, irradiated disk. \textit{The Astrophysical Journal}, \textit{705}(2), 1206.\\

\hl{Charnoz, S., Avice, G., Hyodo, R., Pignatale, F. C., \& Chaussidon, M. (2021). Forming pressure traps at the snow line to isolate isotopic reservoirs in the absence of a planet. \textit{Astronomy \& Astrophysics}, \textit{652}, A35.}\\

Ciesla, F., \& Lauretta, D. (2005). Radial migration and dehydration of phyllosilicates in the solar nebula. \textit{Earth and Planetary Science Letters}, \textit{231}(1-2), 1-8.\\

Cresswell, P., \& Nelson, R. P. (2008). Three-dimensional simulations of multiple protoplanets embedded in a protostellar disc. \textit{Astronomy \& Astrophysics}, \textit{482}(2), 677-690.\\

Cuzzi, J. N., \& Zahnle, K. J. (2004). Material enhancement in protoplanetary nebulae by particle drift through evaporation fronts. \textit{The Astrophysical Journal}, \textit{614}(1), 490.\\

Dai, F., Goldberg, M., Batygin, K., van Saders, J., Chiang, E., Choksi, N., ... \& Winn, J. N. (2024). The prevalence of resonance among young, close-in planets. \textit{The Astronomical Journal}, \textit{168}(6), 239.\\

de Kleer, K., Hughes, E. C., Nimmo, F., Eiler, J., Hofmann, A. E., Luszcz-Cook, S., \& Mandt, K. (2024). Isotopic evidence of long-lived volcanism on Io. \textit{Science}, \textit{384}(6696), 682-687.\\

Drażkowska, J., \& Alibert, Y. (2017). Planetesimal formation starts at the snow line. \textit{Astronomy \& Astrophysics}, \textit{608}, A92.\\

Dubrulle, B., Morfill, G., \& Sterzik, M. (1995). The dust subdisk in the protoplanetary nebula. \textit{icarus}, \textit{114}(2), 237-246.\\

\hl{Dullemond, C. P., Dominik, C., \& Natta, A. (2001). Passive irradiated circumstellar disks with an inner hole. \textit{The Astrophysical Journal}, \textit{560}(2), 957.}\\

Dullemond, C. P., Birnstiel, T., Huang, J., Kurtovic, N. T., Andrews, S. M., Guzmán, V. V., ... \& Ricci, L. (2018). The disk substructures at high angular resolution project (DSHARP). VI. Dust trapping in thin-ringed protoplanetary disks. \textit{The Astrophysical Journal Letters}, \textit{869}(2), L46.\\

Dullemond, C. P., \& Dominik, C. (2005). Dust coagulation in protoplanetary disks: A rapid depletion of small grains. \textit{Astronomy \& Astrophysics}, \textit{434}(3), 971-986.\\

Dullemond, C. P., \& Penzlin, A. B. T. (2018). Dust-driven viscous ring-instability in protoplanetary disks. \textit{Astronomy \& Astrophysics}, \textit{609}, A50.\\

Flock, M., Ruge, J. P., Dzyurkevich, N., Henning, T., Klahr, H., \& Wolf, S. (2015). Gaps, rings, and non-axisymmetric structures in protoplanetary disks-from simulations to alma observations. \textit{Astronomy \& Astrophysics}, \textit{574}, A68.\\

Fuller, J., Luan, J., \& Quataert, E. (2016). Resonance locking as the source of rapid tidal migration in the Jupiter and Saturn moon systems. \textit{Monthly Notices of the Royal Astronomical Society}, \textit{458}(4), 3867-3879.\\

Gao, P., \& Stevenson, D. J. (2013). Nonhydrostatic effects and the determination of icy satellites’ moment of inertia. \textit{Icarus}, \textit{226}(2), 1185-1191.\\

Ghosh, P., \& Lamb, F. K. (1979). Accretion by rotating magnetic neutron stars. II-Radial and vertical structure of the transition zone in disk accretion. \textit{Astrophysical Journal, Part 1, vol. 232, Aug. 15, 1979, p. 259-276.}, \textit{232}, 259-276.\\

Gillon, M., Triaud, A. H., Demory, B. O., Jehin, E., Agol, E., Deck, K. M., ... \& Queloz, D. (2017). Seven temperate terrestrial planets around the nearby ultracool dwarf star TRAPPIST-1. \textit{Nature}, \textit{542}(7642), 456-460.\\

Goldberg, M., \& Batygin, K. (2022). Architectures of compact super-Earth systems shaped by instabilities. \textit{The Astronomical Journal}, \textit{163}(5), 201.\\

Goldreich, P., \& Sciama, D. W. (1965). An explanation of the frequent occurrence of commensurable mean motions in the solar system. \textit{Monthly Notices of the Royal Astronomical Society}, \textit{130}(3), 159-181.\\

Goldreich, P., \& Tremaine, S. (1979). The excitation of density waves at the Lindblad and corotation resonances by an external potential. \textit{Astrophysical Journal}, \textit{233}(3), 857-871.\\

Goldreich, P., \& Tremaine, S. (1980). Disk-satellite interactions. \textit{Astrophysical Journal, Part 1, vol. 241, Oct. 1, 1980, p. 425-441.}, \textit{241}, 425-441.\\

Grasset, O., Dougherty, M. K., Coustenis, A., Bunce, E. J., Erd, C., Titov, D., ... \& Van Hoolst, T. (2013). JUpiter ICy moons Explorer (JUICE): An ESA mission to orbit Ganymede and to characterise the Jupiter system. \textit{Planetary and Space Science}, \textit{78}, 1-21.\\

Gammie, C. F. (1996). Layered accretion in T Tauri disks. \textit{Astrophysical Journal v. 457, p. 355}, \textit{457}, 355.\\

Greenberg, R. (1987). Galilean satellites: Evolutionary paths in deep resonance. \textit{Icarus}, \textit{70}(2), 334-347.\\

Güttler, C., Blum, J., Zsom, A., Ormel, C. W., \& Dullemond, C. P. (2010). The outcome of protoplanetary dust growth: pebbles, boulders, or planetesimals?-I. Mapping the zoo of laboratory collision experiments. \textit{Astronomy \& Astrophysics}, \textit{513}, A56.\\

Hasegawa, Y., \& Pudritz, R. E. (2010). Dead zones as thermal barriers to rapid planetary migration in protoplanetary disks. \textit{The Astrophysical Journal Letters}, \textit{710}(2), L167.\\

Johansen, A., Youdin, A., \& Klahr, H. (2009). Zonal flows and long-lived axisymmetric pressure bumps in magnetorotational turbulence. \textit{The Astrophysical Journal}, \textit{697}(2), 1269.\\

Kivelson, M. G., Khurana, K. K., Russell, C. T., Volwerk, M., Walker, R. J., \& Zimmer, C. (2000). Galileo magnetometer measurements: A stronger case for a subsurface ocean at Europa. \textit{Science}, \textit{289}(5483), 1340-1343.\\

Kivelson, M. G., Khurana, K. K., \& Volwerk, M. (2002). The permanent and inductive magnetic moments of Ganymede. \textit{Icarus}, \textit{157}(2), 507-522.\\

Kleine, T., Budde, G., Burkhardt, C., Kruijer, T. S., Worsham, E. A., Morbidelli, A., \& Nimmo, F. (2020). The non-carbonaceous–carbonaceous meteorite dichotomy. \textit{Space Science Reviews}, \textit{216}, 1-27.\\

Kley, W., \& Nelson, R. P. (2012). Planet-disk interaction and orbital evolution. \textit{Annual Review of Astronomy and Astrophysics}, \textit{50}(1), 211-249.\\

Kretke, K. A., \& Lin, D. N. C. (2007). Grain retention and formation of planetesimals near the snow line in MRI-driven turbulent protoplanetary disks. \textit{The Astrophysical Journal}, \textit{664}(1), L55.\\

Kruijer, T. S., Burkhardt, C., Budde, G., \& Kleine, T. (2017). Age of Jupiter inferred from the distinct genetics and formation times of meteorites. \textit{Proceedings of the National Academy of Sciences}, \textit{114}(26), 6712-6716.\\

Kuskov, O. L., \& Kronrod, V. A. (2001). Core sizes and internal structure of Earth's and Jupiter's satellites. \textit{Icarus}, \textit{151}(2), 204-227.\\

Li, H. F. J. M., Finn, J. M., Lovelace, R. V. E., \& Colgate, S. A. (2000). Rossby wave instability of thin accretion disks. II. Detailed linear theory. \textit{The Astrophysical Journal}, \textit{533}(2), 1023.\\

Lichtenberg, T., Drażkowska, J., Schönbächler, M., Golabek, G. J., \& Hands, T. O. (2021). Bifurcation of planetary building blocks during Solar System formation. \textit{Science}, \textit{371}(6527), 365-370.\\

Lubow, S. H., \& Martin, R. G. (2013). Dead zones in circumplanetary discs as formation sites for regular satellites. \textit{Monthly Notices of the Royal Astronomical Society}, \textit{428}(3), 2668-2673.\\

Luger, R., Sestovic, M., Kruse, E., Grimm, S. L., Demory, B. O., Agol, E., ... \& Queloz, D. (2017). A seven-planet resonant chain in TRAPPIST-1. \textit{Nature Astronomy}, \textit{1}(6), 0129.\\

Lynden-Bell, D., \& Pringle, J. E. (1974). The evolution of viscous discs and the origin of the nebular variables. \textit{Monthly Notices of the Royal Astronomical Society}, \textit{168}(3), 603-637.\\

Madeira, G., Izidoro, A., \& Giuliatti Winter, S. M. (2021). Building the Galilean moons system via pebble accretion and migration: a primordial resonant chain. \textit{Monthly Notices of the Royal Astronomical Society}, \textit{504}(2), 1854-1872.\\

Malhotra, R. (1991). Tidal origin of the Laplace resonance and the resurfacing of Ganymede. \textit{Icarus}, \textit{94}(2), 399-412.\\

Masset, F. S., Morbidelli, A., Crida, A., \& Ferreira, J. (2006). Disk surface density transitions as protoplanet traps. \textit{The Astrophysical Journal}, \textit{642}(1), 478.\\

McEwen, A. S., Belton, M. J. S., Breneman, H. H., Fagents, S. A., Geissler, P., Greeley, R., ... \& Williams, D. A. (2000). Galileo at Io: Results from high-resolution imaging. \textit{Science}, \textit{288}(5469), 1193-1198.\\

McKinnon, W. B. (2023). Setting the stage: formation and earliest evolution of Io. In \textit{Io: A New View of Jupiter’s Moon} (pp. 41-93). Cham: Springer International Publishing.\\

Mills, S., Fabrycky, D. C., Migaszewski, C., Ford, E. B., Petigura, E., \& Isaacson, H. T. (2016, May). Kepler-223: A resonant chain of four transiting, sub-Neptune planets. In \textit{AAS/Division of Dynamical Astronomy Meeting\# 47} (Vol. 47, pp. 101-03).\\

Mohanty, S., \& Shu, F. H. (2008). Magnetocentrifugally driven flows from young stars and disks. VI. Accretion with a multipole stellar field. \textit{The Astrophysical Journal}, \textit{687}(2), 1323.\\

Morbidelli, A., Baillie, K., Batygin, K., Charnoz, S., Guillot, T., Rubie, D. C., \& Kleine, T. (2022). Contemporary formation of early Solar System planetesimals at two distinct radial locations. \textit{Nature Astronomy}, \textit{6}(1), 72-79.\\

Morbidelli, A., Szulágyi, J., Crida, A., Lega, E., Bitsch, B., Tanigawa, T., \& Kanagawa, K. (2014). Meridional circulation of gas into gaps opened by giant planets in three-dimensional low-viscosity disks. \textit{Icarus}, \textit{232}, 266-270.\\

Mosqueira, I., \& Estrada, P. R. (2003). Formation of the regular satellites of giant planets in an extended gaseous nebula I: subnebula model and accretion of satellites. \textit{Icarus}, \textit{163}(1), 198-231.

Mueller, S., \& McKinnon, W. B. (1988). Three-layered models of Ganymede and Callisto: Compositions, structures, and aspects of evolution. \textit{Icarus}, \textit{76}(3), 437-464.\\

\hl{Müller, J., Savvidou, S., \& Bitsch, B. (2021). The water-ice line as a birthplace of planets: implications of a species-dependent dust fragmentation threshold. \textit{Astronomy \& Astrophysics}, \textit{650}, A185.}\\

Murray, C. D., \& Dermott, S. F. (1999). \textit{Solar system dynamics}. Cambridge university press.\\

Nakamoto, T., \& Nakagawa, Y. (1994). Formation, early evolution, and gravitational stability of protoplanetary disks. \textit{Astrophysical Journal, Part 1 (ISSN 0004-637X), vol. 421, no. 2, p. 640-650}, \textit{421}, 640-650.\\

Ogilvie, G. I., \& Lubow, S. H. (2003). Saturation of the corotation resonance in a gaseous disk. \textit{The Astrophysical Journal}, \textit{587}(1), 398.\\

Okuzumi, S., \& Hirose, S. (2011). Modeling magnetorotational turbulence in protoplanetary disks with dead zones. \textit{The Astrophysical Journal}, \textit{742}(2), 65.\\

Ostriker, E. C., \& Shu, F. H. (1995). Magnetocentrifugally driven flows from young stars and disks. IV. The accretion funnel and dead zone. \textit{Astrophysical Journal v. 447, p. 813}, \textit{447}, 813.\\

Paardekooper, S. J., Baruteau, C., \& Kley, W. (2011). A torque formula for non-isothermal Type I planetary migration–II. Effects of diffusion. \textit{Monthly Notices of the Royal Astronomical Society}, \textit{410}(1), 293-303.\\

Paardekooper, S. J., \& Mellema, G. (2006). Halting type I planet migration in non-isothermal disks. \textit{Astronomy \& Astrophysics}, \textit{459}(1), L17-L20.\\

Paardekooper, S. J., \& Mellema, G. (2008). Growing and moving low-mass planets in non-isothermal disks. \textit{Astronomy \& Astrophysics}, \textit{478}(1), 245-266.\\

Paardekooper, S. J., \& Papaloizou, J. C. (2008). On disc protoplanet interactions in a non-barotropic disc with thermal diffusion. \textit{Astronomy \& Astrophysics}, \textit{485}(3), 877-895.\\

Peale, S. J., Cassen, P., \& Reynolds, R. T. (1979). Melting of Io by tidal dissipation. \textit{Science}, \textit{203}(4383), 892-894.\\

Peale, S. J. (1999). Origin and evolution of the natural satellites. \textit{Annual Review of Astronomy and Astrophysics}, \textit{37}(1), 533-602.\\

Peale, S. J., \& Lee, M. H. (2002). A primordial origin of the Laplace relation among the Galilean satellites. \textit{Science}, \textit{298}(5593), 593-597.\\

Pichierri, G., Bitsch, B., \& Lega, E. (2023). A recipe for orbital eccentricity damping in the type-I regime for low-viscosity 2D discs. \textit{Astronomy \& Astrophysics}, \textit{670}, A148.\\

Pichierri, G., Bitsch, B., \& Lega, E. (2024b). A recipe for eccentricity and inclination damping for partial-gap opening planets in 3D disks. \textit{The Astrophysical Journal}, \textit{967}(2), 111.\\

Pichierri, G., Morbidelli, A., Batygin, K., \& Brasser, R. (2024a). The formation of the TRAPPIST-1 system in two steps during the recession of the disk inner edge. \textit{Nature Astronomy}, 1-8.\\

Pinilla, P., Flock, M., de Juan Ovelar, M., \& Birnstiel, T. (2016). Can dead zones create structures like a transition disk?. \textit{Astronomy \& Astrophysics}, \textit{596}, A81.\\

Rein, H., \& Liu, S. F. (2012). REBOUND: an open-source multi-purpose N-body code for collisional dynamics. \textit{Astronomy \& Astrophysics}, \textit{537}, A128.\\

Rein, H., \& Tamayo, D. (2015). WHFAST: a fast and unbiased implementation of a symplectic Wisdom–Holman integrator for long-term gravitational simulations. \textit{Monthly Notices of the Royal Astronomical Society}, \textit{452}(1), 376-388.\\

Riols, A., \& Lesur, G. (2018). Dust settling and rings in the outer regions of protoplanetary discs subject to ambipolar diffusion. \textit{Astronomy \& Astrophysics}, \textit{617}, A117.\\

Ros, K., \& Johansen, A. (2013). Ice condensation as a planet formation mechanism. \textit{Astronomy \& Astrophysics}, \textit{552}, A137.\\

Sasaki, T., Stewart, G. R., \& Ida, S. (2010). Origin of the different architectures of the Jovian and Saturnian satellite systems. \textit{The Astrophysical Journal}, \textit{714}(2), 1052.\\

\hl{Schneeberger, A., \& Mousis, O. (2025). Impact of Jupiter’s Heating and Self-shadowing on the Jovian Circumplanetary Disk Structure. \textit{The Planetary Science Journal}, \textit{6}(1), 23.}\\

Schubert, G., Anderson, J. D., Spohn, T., \& McKinnon, W. B. (2004). Interior composition, structure and dynamics of the Galilean satellites. \textit{Jupiter: The planet, satellites and magnetosphere}, \textit{1}, 281-306.\\

Shakura, N. I., \& Sunyaev, R. A. (1973). Black holes in binary systems. Observational appearance. \textit{Astronomy and Astrophysics, Vol. 24, p. 337-355}, \textit{24}, 337-355.\\

Shibaike, Y., Ormel, C. W., Ida, S., Okuzumi, S., \& Sasaki, T. (2019). The Galilean satellites formed slowly from pebbles. \textit{The Astrophysical Journal}, \textit{885}(1), 79.\\

Showman, A. P., \& Malhotra, R. (1997). Tidal evolution into the Laplace resonance and the resurfacing of Ganymede. \textit{Icarus}, \textit{127}(1), 93-111.\\

Sohl, F., Spohn, T., Breuer, D., \& Nagel, K. (2002). Implications from Galileo observations on the interior structure and chemistry of the Galilean satellites. \textit{Icarus}, \textit{157}(1), 104-119.\\

Stevenson, D. J., \& Lunine, J. I. (1988). Rapid formation of Jupiter by diffusive redistribution of water vapor in the solar nebula. \textit{Icarus}, \textit{75}(1), 146-155.\\

Szulágyi, J., Binkert, F., \& Surville, C. (2022). Meridional circulation of dust and gas in the circumstellar disk: Delivery of solids onto the circumplanetary region. \textit{The Astrophysical Journal}, \textit{924}(1), 1.\\

Tabone, B., Rosotti, G. P., Cridland, A. J., Armitage, P. J., \& Lodato, G. (2022). Secular evolution of MHD wind-driven discs: analytical solutions in the expanded $\alpha$-framework. \textit{Monthly Notices of the Royal Astronomical Society}, \textit{512}(2), 2290-2309.\\

Takahashi, S. Z., \& Inutsuka, S. I. (2014). Two-component secular gravitational instability in a protoplanetary disk: a possible mechanism for creating ring-like structures. \textit{The Astrophysical Journal}, \textit{794}(1), 55.\\

Tamayo, D., Rein, H., Shi, P., \& Hernandez, D. M. (2020). REBOUNDx: a library for adding conservative and dissipative forces to otherwise symplectic N-body integrations. \textit{Monthly Notices of the Royal Astronomical Society}, \textit{491}(2), 2885-2901.\\

Tanaka, H., Takeuchi, T., \& Ward, W. R. (2002). Three-dimensional interaction between a planet and an isothermal gaseous disk. I. Corotation and Lindblad torques and planet migration. \textit{The Astrophysical Journal}, \textit{565}(2), 1257.\\

Tanaka, H., \& Ward, W. R. (2004). Three-dimensional interaction between a planet and an isothermal gaseous disk. II. Eccentricity waves and bending waves. \textit{The Astrophysical Journal}, \textit{602}(1), 388.\\

Tanigawa, T., Ohtsuki, K., \& Machida, M. N. (2012). Distribution of accreting gas and angular momentum onto circumplanetary disks. \textit{The Astrophysical Journal}, \textit{747}(1), 47.\\

Teague, R., Bae, J., Bergin, E. A., Birnstiel, T., \& Foreman-Mackey, D. (2018). A kinematical detection of two embedded Jupiter-mass planets in HD 163296. \textit{The Astrophysical Journal Letters}, \textit{860}(1), L12.\\

Teague, R., Bae, J., \& Bergin, E. A. (2019). Meridional flows in the disk around a young star. \textit{Nature}, \textit{574}(7778), 378-381.\\

\hl{Tissot, F. L., Burkhardt, C., Kuznetsova, A., Pack, A., Schiller, M., Spitzer, F., ... \& Yap, T. E. (2025). Infall and Disk Processes–the Message from Meteorites. \textit{Space Science Reviews}, \textit{221}(7), 1-50.}\\

Tripathi, A., Andrews, S. M., Birnstiel, T., \& Wilner, D. J. (2017). A millimeter continuum size–luminosity relationship for protoplanetary disks. \textit{The Astrophysical Journal}, \textit{845}(1), 44.\\

Wang, H., Weiss, B. P., Bai, X. N., Downey, B. G., Wang, J., Wang, J., ... \& Zucolotto, M. E. (2017). Lifetime of the solar nebula constrained by meteorite paleomagnetism. \textit{Science}, \textit{355}(6325), 623-627.\\

Ward, W. R. (1997). Protoplanet migration by nebula tides. \textit{Icarus}, \textit{126}(2), 261-281.\\

Warren, P. H. (2011). Stable-isotopic anomalies and the accretionary assemblage of the Earth and Mars: A subordinate role for carbonaceous chondrites. \textit{Earth and Planetary Science Letters}, \textit{311}(1-2), 93-100.\\

Weiss, B. P., Bai, X. N., \& Fu, R. R. (2021). History of the solar nebula from meteorite paleomagnetism. \textit{Science Advances}, \textit{7}(1), eaba5967.\\

Weiss, L. M., Marcy, G. W., Petigura, E. A., Fulton, B. J., Howard, A. W., Winn, J. N., ... \& Cargile, P. A. (2018). The California-Kepler survey. V. Peas in a pod: Planets in a Kepler multi-planet system are similar in size and regularly spaced. \textit{The Astronomical Journal}, \textit{155}(1), 48.\\

Williams, J. P., \& Cieza, L. A. (2011). Protoplanetary disks and their evolution. \textit{Annual Review of Astronomy and Astrophysics}, \textit{49}(1), 67-117.\\

Yang, C. C., Mac Low, M. M., \& Johansen, A. (2018). Diffusion and concentration of solids in the dead zone of a protoplanetary disk. \textit{The Astrophysical Journal}, \textit{868}(1), 27.\\

Yap, T. E., \& Batygin, K. (2024). Dust-gas coupling in turbulence-and MHD wind-driven protoplanetary disks: Implications for rocky planet formation. \textit{Icarus}, \textit{417}, 116085.\\

Yap, T. E., \& Stevenson, D. J. (2023, December). Evaluating the Conversion of CO to CH4 in the Jovian Circumplanetary Disk as a Source of Abundant Water Ice in Ganymede and Callisto. In \textit{AGU Fall Meeting Abstracts} (Vol. 2023, No. 3171, pp. P33C-3171).\\

Yap, T. E., \& Tissot, F. L. (2023). The NC-CC dichotomy explained by significant addition of CAI-like dust to the Bulk Molecular Cloud (BMC) composition. \textit{Icarus}, \textit{405}, 115680.\\

Yoder, C. F. (1979). How tidal heating in Io drives the Galilean orbital resonance locks. \textit{Nature}, \textit{279}(5716), 767-770.\\

Yoder, C. F., \& Peale, S. J. (1981). The tides of Io. \textit{Icarus}, \textit{47}(1), 1-35.\\

\section*{Supplementary figures}
\indent \textit{See below:}

\begin{figure}
\scalebox{0.55}{\includegraphics{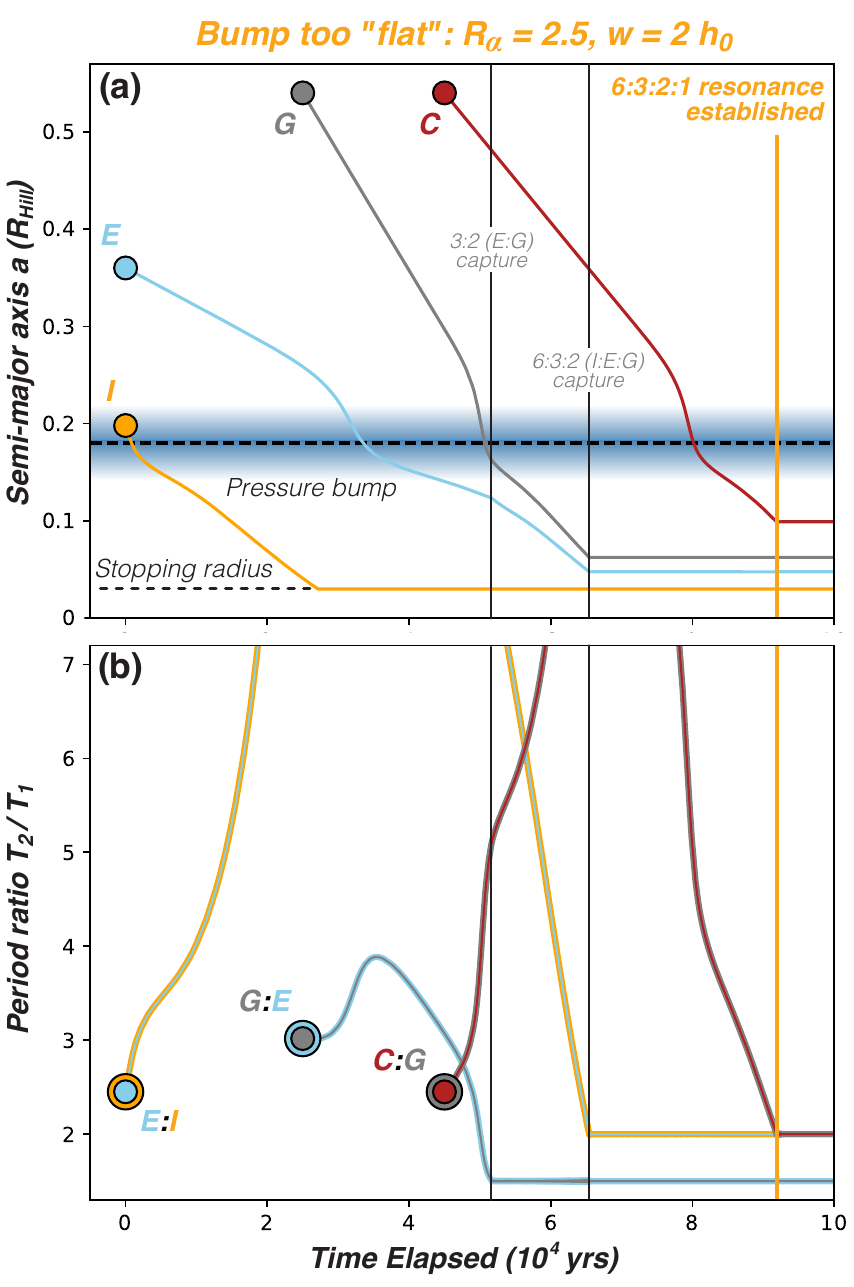}}
\caption{\textbf{; S1. Simulation results for $R_{stop}= 0.03 R_{Hill}$ and $r_0 = 0.18 R_{Hill}$ , with $R_{\alpha} = 2.5$ and $w = 2h_0$. }Panels indicate the \textbf{(a)} semi-major axes of the moons and \textbf{(b)} their outer-inner period ratios. Key resonant captures are denoted by vertical lines. Here, the bump is too ``flat" (\textit{i.e.,} short/wide; $\Delta h/w \lesssim 0.45$) to function as a migration trap. As such, all four moons individually make it past the bump, establishing a 6:3:2:1 resonance interior to it. See \textbf{Section 4.2} for discussion, and \textbf{Fig. 3} for all $R_{\alpha}-w$ pairs explored that correspond to this final outcome.}
\label{fig:Figure S1}
\end{figure}

\begin{figure*} 
\scalebox{0.55}{\includegraphics{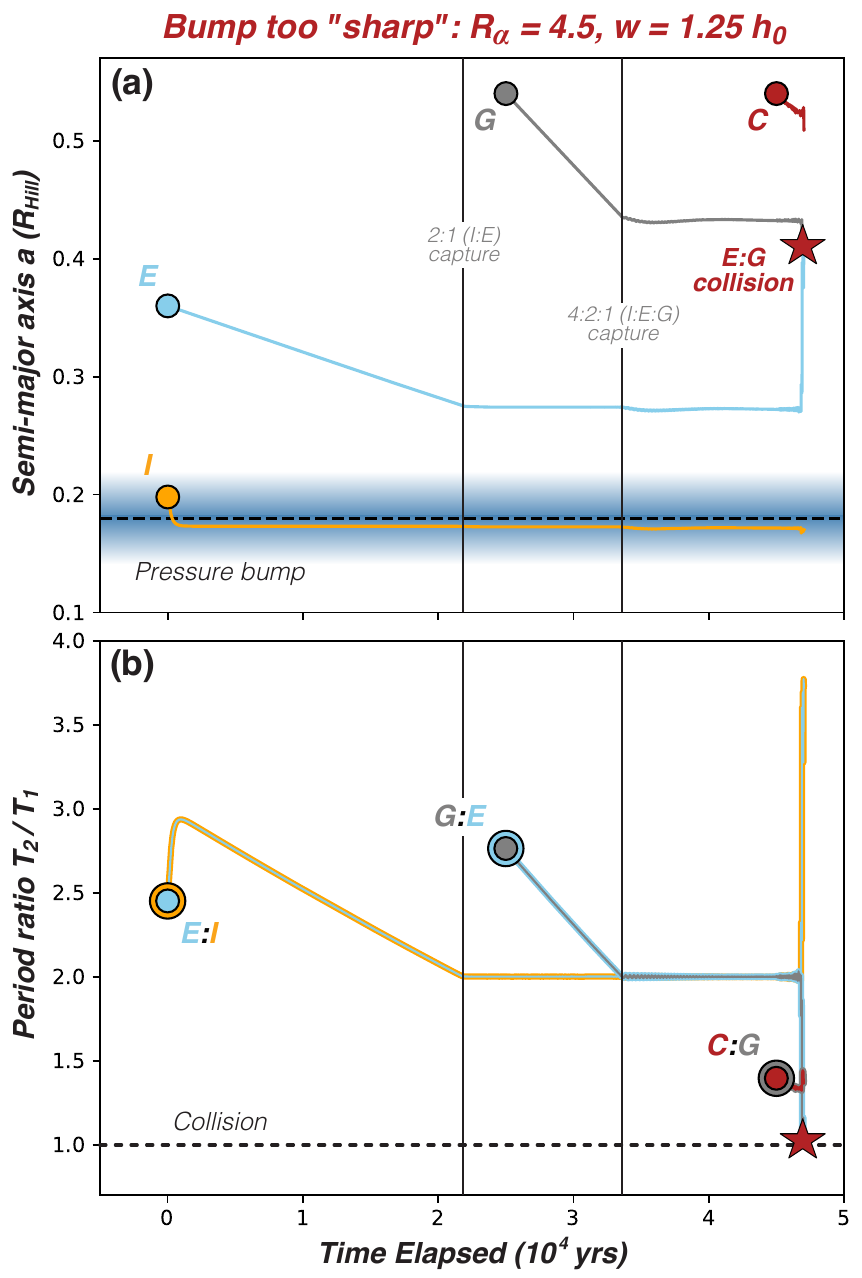}}
\caption{\textbf{; S2. Simulation results for $R_{stop}= 0.03 R_{Hill}$ and $r_0 = 0.18 R_{Hill}$ , with $R_{\alpha} = 4.5$ and $w = 1.25h_0$. }Panels indicate the \textbf{(a)} semi-major axes of the moons and \textbf{(b)} their outer-inner period ratios. Key resonant captures are denoted by vertical lines. Here, the bump is too ``sharp" (\textit{i.e.,} tall/thin; $\Delta h/w \gtrsim 0.6$) to allow for the trapped moon (\textit{i.e.,} \textit{I}) to be ``pushed" across once resonance is established, even with \textit{G}. Eccentricities are ``pumped" till a collision between \textit{E} and \textit{G} takes place, terminating the simulation. See \textbf{Section 4.2} for discussion, and \textbf{Fig. 3} for all $R_{\alpha}-w$ pairs explored that correspond to resonant pile-ups at the bump, and eventual instability.}
\label{fig:Figure S2}
\end{figure*}

\begin{figure*} 
\scalebox{0.55}{\includegraphics{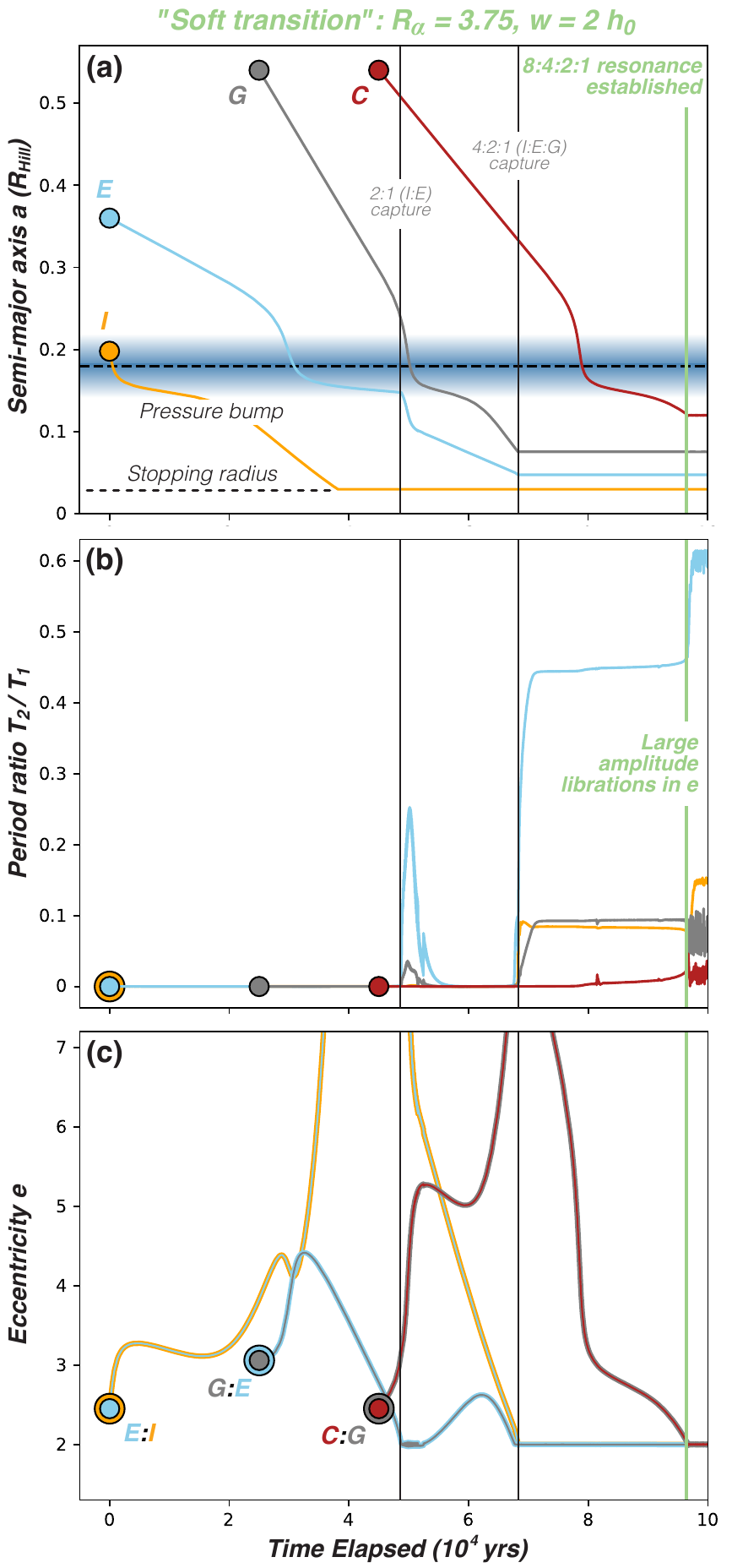}}
\caption{\textbf{; S3. Simulation results for $R_{stop}= 0.03 R_{Hill}$ and $r_0 = 0.18 R_{Hill}$ , with $R_{\alpha} = 3.75$ and $w = 2h_0$. }Panels indicate the \textbf{(a)} semi-major axes and \textbf{(b)} eccentricities of the moons, as well as \textbf{(c)} their outer-inner period ratios. Key resonant captures are denoted by vertical lines. This simulation, culminating in a 8:4:2:1 resonance, is an example of those constituting the ``soft" transition between the regimes wherein (i) the intended result (\textit{i.e.,} 4:2:1 resonance between \textit{I}, \textit{E}, and \textit{G}; \textit{C} trapped at bump) and (ii) a 6:3:2:1 resonance between the four moons are achieved (see \textbf{Fig. 3}). Variations on simulation outcomes at the ``soft" transition (at $\Delta h/w \sim 0.45$) include 12:6:3:2, 16:8:6:3, and 9:6:4:2 resonances. See \textbf{Section 4.2} for discussion.}
\label{fig:Figure S3}
\end{figure*}

\begin{figure*} 
\scalebox{0.5}{\includegraphics{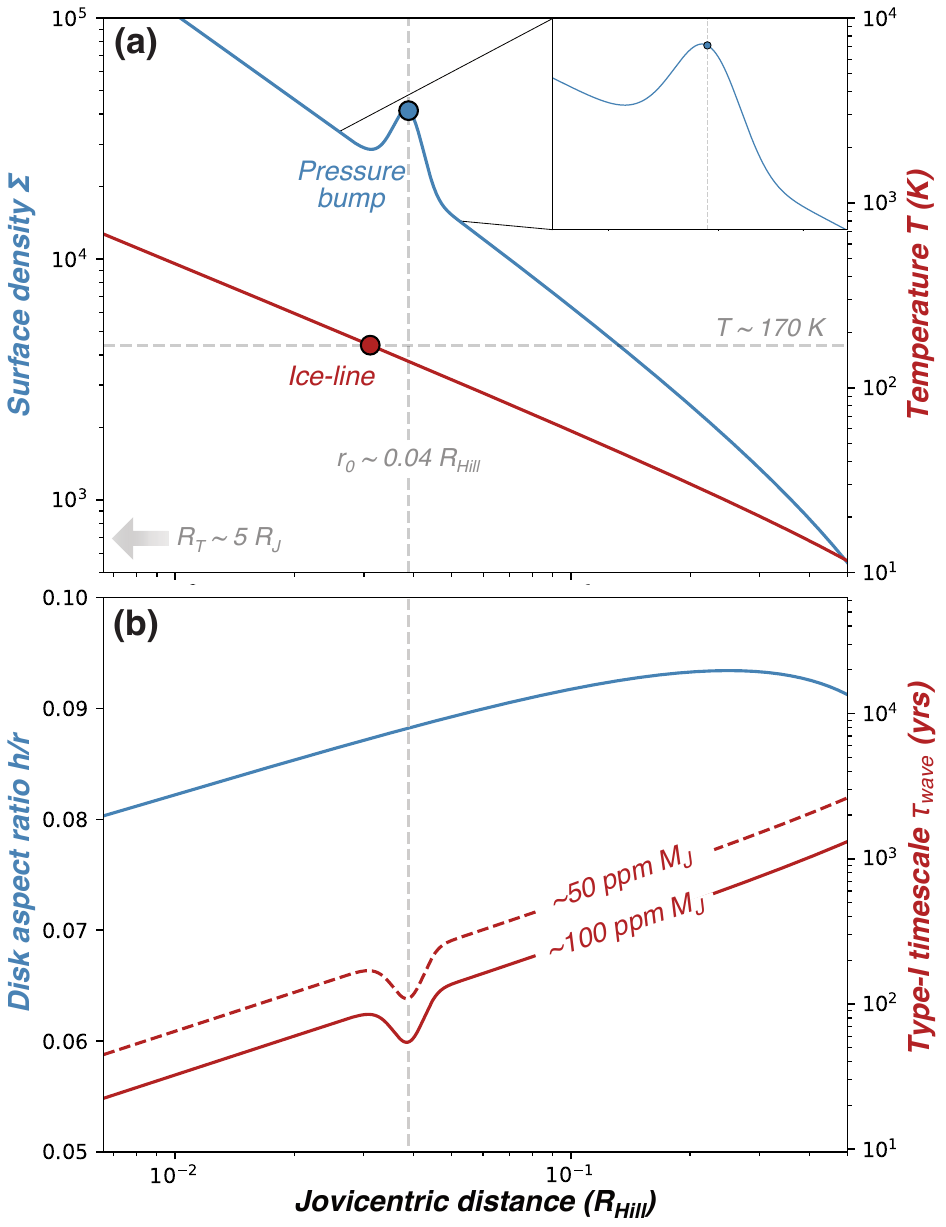}}
\caption{\textbf{; S4. Plots of (a) the surface density and temperature profile of the disk, as well as (b) the disk aspect ratio and the characteristic Type-I timescale for two hypothetical moons (one with $\sim 50$ ppm and the other $\sim 100$ ppm of Jupiter's mass), where $R_{\alpha} = 2$ and $w = h_0$.} Inset in (a) depicts a close-up view of the pressure bump, just beyond (\textit{i.e.,} 1.25 times) the ice-line ($r_0\sim 0.04 R_{Hill}$). As discussed in \textbf{Section 2.3}, larger bodies migrate and, have their eccentricities damped, faster. Simulation results corresponding to these $\Sigma(r)$ and $h(r)/r$ profiles are discussed in \textbf{Section 4.3}, and presented in \textbf{Fig. 4}. }
\label{fig:Figure S4}
\end{figure*}

\end{document}